\documentclass[12pt]{article}
\begin{document}
\renewcommand{\theequation}{\thesection.\arabic{equation}}

\title{\bf SUPERFLUID TO NORMAL PHASE TRANSITION AND
EXTREME REGULARITY OF SUPERDEFORMED BANDS}

\author{I. M. Pavlichenkov}
\maketitle

\vspace{3cm}
\centerline{\Large Abstract}
\vspace{10mm}

We derive the exact semiclassical expression for the second inertial
parameter $\cal B$ for the superfluid and normal phases. Interpolation
between these limiting values shows that the function ${\cal B}(I)$
changes sign at the spin $I_c$, which is critical for a rotational
spectrum. The quantity $\cal B$ turns out to be a sensitive measure
of the change in static pairing correlations. The superfluid-to-normal
transition reveals itself in the specific variation of the ratio
${\cal B}/{\cal A}$ versus spin $I$ with the plateau characteristic
of the normal phase. We find this dependence to be universal for
normal deformed and superdeformed bands. The long plateau with
a small  value ${\cal B}/{\cal A}\sim A^{-8/3}$ explains the
extreme regularity of superdeformed bands.

\newpage

\setcounter{equation}{0}
\section{Introduction}

{\it Recently phase transitions in mesoscopic systems have been a
subject of intense discussions in nuclear and solid-state physics.
Isai Isidorovitch Gurevich appears to be the first to bring up
(in 1939) the concept
of the temperature phase transition to nuclear studies \cite{Gur}. His
prediction was based on the observation that the level density of
the resonant states formed by a thermal neutron capture is an
unsteady function of the atomic mass number with a maximum in
the rare-earth nuclei.}

The problem of the rotation-induced transition from the superfluid
to the normal phase in nuclei has been a foremost theme in
high-spin spectroscopy since Mottelson and Valatin
\cite{MV} predicted a pairing collapse in rapidly rotating nuclei.
This effect can be understood by an analogy with a superconductor
in a magnetic field. In a deformed nucleus the Cooper
pair is formed by two nucleons with opposite single-particle
angular momentum projections $\pm m$. Being time noninvariant
(as a magnetic field) the Coriolis force in a rotating nucleus acts
on both nucleons in opposite directions and tries to decrease the
spatial overlap of these time reversal orbits. The Coriolis force
increases proportionally with the spin of a band.
Therefore, at some critical spin one may expect that all pairs
are broken and pairing correlations disappear completely.
The phenomenon can be observed by the crossing of the ground
state superfluid band with the band based on the normal state.
Thus, the rigid body moment of inertia corresponding to the
second band appears to be the obvious signature of the pairing
phase transition.

However, this regime of the transition to the normal phase is not
realized in nuclei because they are finite systems with a shell
structure and a small number of nucleons involved in pairing
correlations. The Coriolis force in a rotating nucleus is proportional
to the value of the single-particle angular momentum $j$ of a
nucleon. Thus, the Coriolis antipairing effect is strongest for
nucleons occupying the states of subshells with the largest $j$. In
the vicinity of the Fermi surface, these so-called intruder orbitals
arise from the $j={\cal N}+1/2$ subshell, where $\cal N$ is
the principal quantum number of the shell above. Therefore
they are distinguished from other states of the unfilled shell by the
parity. At normal deformations (ND), the intruder states retain their
$j$ quantum number, while at superdeformations (SD) the $j$-subshell
notation becomes less appropriate due to mixing. Initially the
Coriolis force breaks only the one Cooper pair that
belongs to the intruder orbitals, whereas the rest of the pairs stay
correlated. The band built on such two-quasiparticle excitation
(the rotationally-aligned band) is characterized, due to the blocking
effect, by appreciably smaller pairing correlations than in the
ground state band. Having the largest moment of inertia, the former
crosses the ground state band and becomes yrast. The relevant
phenomenon is backbending. Subsequent breaking of correlated
pairs and their alignment makes the internal structure of the yrast
band non-homogeneous and the transition to the
normal phase configuration dependent.

The standard definition of the phase transition is based on the
mean-field approximation in which different phases are distinguished
by the order parameter, i. e., the static pairing gap $\Delta$. However
the mean field approach to the nuclear pairing correlations faces a
fundamental problem of the quantum fluctuations, which become
quite strong for finite systems. The fluctuating part $\delta\Delta$
(dynamical pairing correlations) of the order parameter is comparable
with $\Delta$ in the transition region. The fluctuations smear out a
sharp phase transition and make it a difficult issue to find the
experimental signature of the superfluid to normal phase transition
in rotational bands. For example, the dependence of the kinematic
$\Im^{(1)}$ or dynamical $\Im^{(2)}$ moments of inertia on the spin $I$
is not a definite indicator of the phase transition. Experimental
evidence of the pairing phase transitions has been discussed usually
in terms of the relative excitation spectrum. As shown in
Ref.~\cite{Shi2} the disappearance of static pairing leads to a
change of the excitation spectrum, from the quasiparticle to the
particle-hole spectrum. Unfortunately, the application of this criterion
to normal deformed (ND) bands \cite{Shi2,Gar,Oliv} shows that this
method is not free from ambiguities.

Meanwhile, it is well known that the change in the
system internal structure manifests itself in the modification of its
collective excitation. The examples for finite quantum systems are
numerous. A classical one is the transition from deformed to
spherical nuclei. In this case, the rotational-vibrational spectrum
transforms into a pure vibrational one. The study of the
bifurcations in rotational spectra \cite{Pav} shows an intimate
connection between internal and rotational motion. For example,
the angular momentum
alignment in a band (the change of a coupling scheme) is observed
as an increase of the energy signature splitting \cite{Pav1}.

The transition we study is more delicate. Consider
the simplest rotational sequence with the parity and the signature
$(\pi\alpha)=(+0)$. The relevant energy spectrum can be parametrized
as follows:
\begin{equation}
E(I)={\cal A}I(I+1) + {\cal B}I^2(I+1)^2 ,
\label{exp}
\end{equation}
where ${\cal A}=\hbar^2/2\Im^{(1)}$ and ${\cal B}$ are the first two
inertial parameters. The spectrum
(\ref{exp}) undergoes a noticeable modification if, for example, the
second inertial parameter changes sign at some spin $I_c$. For $I<I_c$
the spectrum is compressed relative to the rigid rotor spectrum
because ${\cal B}$ is negative for the low-$I$ states. However, for
$I>I_c$ the spectrum becomes extended. The effect can be visualized
by using the $I$ dependence of the ratio ${\cal B}/{\cal A}$. The main
objective of the present paper is to study this dependence.

The parameters ${\cal A}$ and ${\cal B}$ are determined by the
$\gamma$-ray transition energies $E_{\gamma}(I)=E(I+2)-E(I)$ as
follows:
\vspace{-2mm}
$$
{\cal A}(I)=\frac{1}{4(2I\!+\!5)}\!\left[\frac{I^2\!+
\!7I\!+\!13}{2I+3}E_{\gamma}(I)\! -\! \frac{I^2\!+\!3I\!+
  \!3}{2I+7}E_{\gamma}(I\!+\!2)\right],
$$
\begin{equation}
{\cal B}(I) = \frac{1}{8(2I+5)}\left[ \frac{E_{\gamma}(I+2)}{2I+7}-
\frac{E_{\gamma}(I)}{2I+3}\right].
\label{para}
\end{equation}

The coefficient ${\cal B}$ characterizes the nonadiabatic
properties of a  band and is very sensitive to its internal structure.
It also realizes the relationship between kinematic and dynamic
moments of inertia. Using the well-known expressions for these
quantities (see, for example, Ref.~\cite{Fir}) and the last formula
(\ref{para}) we get
\begin{equation}
{\cal B} = \frac{\hbar^2}{2(2I+3)(2I+7)}\left[ \frac{1}{\Im^{(2)}}-
\frac{2I}{(2I+5)\Im^{(1)}}\right].
\label{moi}
\end{equation}
Thus, the parameter ${\cal B}(I)$ is proportional to the difference
$\Im^{(1)}-\Im^{(2)}$ in the high-$I$ limit. The ratio ${\cal B}/{\cal A}$
also determines the convergence radius of the rotational energy
expansion in terms of $I(I+1)$ \cite{B/M}. Faster convergence is
obtained with the Harris formula
\begin{equation}
E(\omega)= E_0 + \frac{1}{2}\alpha\omega^2 +
        \frac{3}{4}\beta\omega^4 + ...,
\label{Hexp}
\end{equation}
which is based on the fourth-order cranking expansion
\begin{equation}
\alpha = \frac{1}{\omega}{\rm tr}(j_x\rho^{(1)}),  \hspace{5mm}
 \beta = \frac{1}{\omega^3}{\rm tr}(j_x\rho^{(3)}),
\label{cranc}
\end{equation}
where $\rho^{(n)}$ is the $n$th correction to the nucleus density
matrix; $j_x$ is the projection of the single-particle angular
momentum operator onto the rotational axis $x$, which is
perpendicular to the symmetry axis $z$; and $\omega$ is the
rotational frequency. The latter depends on the system angular
momentum and is determined by
\begin{equation}
\hbar\sqrt{I(I+1)}=\alpha\omega +\beta\omega^3 + ...\ .
\label{spin}
\end{equation}
It follows from Eqs.\ (\ref{exp}), (\ref{Hexp}), and (\ref{spin}) that
\begin{equation}
\alpha=\frac{\hbar^2}{2{\cal A}},\hspace{5mm}
     \beta=-\frac{\hbar^4{\cal B}}{4{\cal A}^4}.
\label{cone}
\end{equation}

The problem of the microscopic calculation of the parameter $\cal B$
for ND nuclei has attracted considerable attention (see the review 
article \cite{Mih} and references therein). It has been shown that this 
quantity receives the contributions from four types of nonadiabatic 
effects:

(i) perturbation of quasiparticle motion by rotation (quasiparticle
alignment),

(ii) attenuation of pairing correlations by the Coriolis force (Coriolis
antipairing effect),

(iii) a change in the deformation of nuclear self-consistent field
(centrifugal stretching effect),

(iv) vibration-rotation interaction. \\
The first attempt to estimate $\cal B$ was made by the author
together with Grin' \cite{GP}. A Green's function formulation of the
Hartree-Fock-Bogolubov (HFB) method was used to find the expansion
(\ref{Hexp}) for the axially deformed oscillator potential as the
self-consistent field. It was shown that the first and the second effects
yield ${\cal B}/{\cal A}\sim A^{-4/3}$ while the centrifugal stretching
contribution is $A^{2/3}$ times smaller for well deformed
nuclei.\footnote{We use the
estimation ${\cal A}\sim\varepsilon_FA^{-5/3}$,
where $\varepsilon_F$ is the Fermi energy and $A$ is the mass
number.} In the subsequent work \cite{Pav2}, the author found that
the vibration-rotation contribution to the parameter ${\cal B}$
accounts for the same $A^{2/3}$ fraction of the main effects.
These results were confirmed by the calculations of Marshalek
\cite{Mar} with the more realistic Nilsson potential.

Thus, the first two effects are dominant for well deformed nuclei.
The quasiparticle alignment depends strongly on
pairing correlations because the pairing force tries to bind
pairs of particles in time reversal states, reducing the ability
of nucleons to carry an angular momentum. Therefore, the
parameter ${\cal B}$ is very sensitive to the variation of the
pairing correlations along a band.

One of the amazing features of SD bands is the extreme regularity
of their rotational spectra. To demonstrate this feature, the
rotational spectra of different axial systems are compared in Fig. 1
with their rigid rotor counterparts. The comparison shows that the SD
band $^{194}$Pb(1) is more regular than the ND band of $^{238}$U
and even the band of the simplest H$_2$ molecule \cite{Dab}. Having
the ratio ${\cal B}/{\cal A}\sim 10^{-5}$, the band $^{194}$Pb(1) is
not a champion among SD bands. For $^{152}$Dy(1), the ratio is of
the order $10^{-6}$ and this is 1000 times smaller than the above
estimation ${\cal B}/{\cal A}\sim A^{-4/3}$. Thus, an SD nucleus is
the best quantum rotor known in nature. Although numerous theoretical
calculations (see, e.g., \cite{Sat,Dob,Hen,Yan}) successfully
reproduce the measured intraband $\gamma$-ray energies, the
underlying microscopic mechanism of this phenomenon has yet to be
well understood.

In this paper we will reveal an interconnection between
the extreme regularity and the transition from the superfluid to the
normal phase. The key to our theoretical approach lies in the
calculation of the second inertial parameter. Compared to
ND bands, there are two features of the pairing correlations
in SD ones which prevent us from using the results of
previous theoretical calculations of the parameter ${\cal B}$
for superdeformation. First, due to the large shell gap stabilizing
the SD minimum, the static pairing field $\Delta$ is small and can
be commensurate with its fluctuation $\delta\Delta$. Qualitative
conclusion concerning the role of the static and dynamic pairing in
SD bands is presented in Ref.~\cite{Shi}. Second, since intruder
single-particle states, which are unavailable at normal deformations,
appear near the Fermi surface in the case of superdeformations,
it is necessary to go beyond the commonly used monopole pairing
force \cite{HN}. The gauge invariant pairing interaction expands the
correlation space and stabilizes the pairing field. The coordinate
dependent (nonuniform) pairing is also crucial for conservation of a
nucleon current in a rotating nucleus \cite{Mig}.

To avoid calculation of the parameter ${\cal B}$ in the
transition region, where pairing fluctuations play an important
role, an interpolation between the values ${\cal B}_s$ and
${\cal B}_n$ is used. The former is associated with the superfluid
phase (where $\Delta\gg\delta\Delta$) and the latter is related to the
normal one ($\Delta=0$). Thus, pairing fluctuations are unessential
for these regions and we can use the mean-field approach.
In the calculation of  ${\cal B}_s$, the nonuniform
pairing induced by rotation is taken into account by using the method
of Ref.~\cite{GP}. It should be noted that the quantity ${\cal B}_s$
found in the cited work is inapplicable for superdeformation.

The paper is organized as follows: In Section \ref{sec2}, the basic
equations of the cranked HFB theory are presented in the framework
of the Green's function method. The spinor form of the Gor'kov
equations is used to simplify calculations in the higher orders of the
perturbation theory. In Section \ref{sec3}, the exact expression for
the second inertial parameter in the superfluid phase is derived with
this technique by applying a semiclassical approximation. The result
is valid for an arbitrary nuclear mean field. The exact analytical
expression for ${\cal B}_s$ is obtained in Section \ref{sec4} in an
axially deformed oscillator potential. In this section we also consider
some limiting cases for this quantity. Of special interest is the
limit of noninteracting nucleons. It is shown that the relevant
parameter ${\cal B}_n$ is positive and smaller than ${\cal B}_s$. The
comparison with available experimental data for SD and ND bands is
presented in Section \ref{sec5}. Section \ref{sec6} concludes and
summarizes the paper. The preliminary results of the present work
have been published in Refs. \cite{Pav3,Pav4}.

\setcounter{equation}{0}
\section{Green's function formalism in the cranking Hartree-Fock-Bogolubov
method}
\label{sec2}

\subsection{Cranked Gor'kov equations}

Our consideration is based on the shell-model Hamiltonian
consisting of the cranked one body term
\begin{equation}
h_\omega({\bf r})=-\frac{{\bf p}^2}{2M} +U({\bf r})-{\bf{\omega\cdot\ell}},
\hspace{5mm} {\bf{\omega}}\{\omega,0,0\}
\label{mfield}
\end{equation}
(where ${\bf p}$ and $M$ are the impulse and the mass of a nucleon
respectively), and the residual short-range interaction, which is
specified by the two body delta-interaction
\begin{equation}
v({\bf r},{\bf r'})=-g\delta({\bf r}-{\bf r'}),  \hspace{5mm} g>0.
\label{int}
\end{equation}
For simplicity, we neglect the spin in the cranking term and consider
only the orbital part ${\bf\ell}$ of the angular momentum ${\bf j}$.
We will also neglect a weak dependence of the self-consistent
deformed potential $U$ on rotation (centrifugal stretching effect).

In the coordinate representation, the
Gor'kov equations \cite{Gor} have the form
$$
[\varepsilon-h_\omega({\bf r})+\varepsilon_F]G({\bf r},{\bf r'},\varepsilon) +
\tilde\Delta({\bf r})F^+({\bf r},{\bf r'},\varepsilon)=\delta({\bf r}-{\bf r'}),
$$
$$
[\varepsilon+h_\omega^+({\bf r})-\varepsilon_F]F^+({\bf r},{\bf r'},\varepsilon) +
\tilde\Delta^+({\bf r})G({\bf r},{\bf r'},\varepsilon)=0,
$$
\begin{equation}
\tilde\Delta^*({\bf r})=
   g\oint\limits_C\frac{d\varepsilon}{2\pi i}F^+({\bf r},{\bf r},\varepsilon).
\label{gork}
\end{equation}
The functions $G({\bf r},{\bf r'},\varepsilon)$ and
$F^+({\bf r},{\bf r'},\varepsilon)$ are the Fourier transforms of the
Green's functions
$$
G({\bf r},{\bf r'},t-t')=-i(\Phi_N|T\{\psi({\bf r},t)\psi^+({\bf r'},t)\}|\Phi_N),
$$
\begin{equation}
F^+({\bf r},{\bf r'},t-t')=-i(\Phi_{N+2}|T\{\psi^+({\bf r},t)\psi^+({\bf r'},t)\}
      |\Phi_N)e^{-2i\varepsilon_Ft},
\label{gfk}
\end{equation}
where $\Phi_N$ and $\Phi_{N+2}$ are the eigenfunctions of the ground
state of a system of $N$ and $N+2$ interacting particles, $\psi^+$ and
$\psi$ are creation and annihilation operators in the Heisenberg
representation, $T$ is the time ordering operator, and
$\varepsilon_F$ is the Fermi energy (the chemical potential of the
system). The contour $C$ consists of the real axis and an infinite
semicircle in the upper half-plane.

\subsection{Properties}

In obtaining Eqs. (\ref{gork}) the particle number nonconserving
approximation has been used. In a spirit of the
mean-field approach we neglect a difference between the functions
$\Phi_N$ and $\Phi_{N+2}$. Thus, the Gor'kov equations describe
a system with the broken gauge symmetry associated with the particle
number. However, the average particle number is fixed. This is
achieved by adding the term $-\varepsilon_F\hat N$ to the Hamiltonian.
The Lagrange multiplier $\varepsilon_F$ is determined by the
equation
\begin{equation}
N=\int d{\bf r}\oint\limits_C\frac{d\varepsilon}{2\pi i}G({\bf r},{\bf r},\varepsilon).
\label{numb}
\end{equation}

Equations (\ref{gork}) are also noninvariant with respect to the more general
gauge transformation (the local Gallileian transformation \cite{Bel})
\begin{equation}
\psi({\bf r},t) \to \psi({\bf r},t)e^{i\phi({\bf r})},
\label{grad}
\end{equation}
where $\phi({\bf r})$ is an arbitrary function of the space coordinates.
The quickest way to show this is to introduce the vector potential
${\bf A}=[{\bf\omega r}]/2$ that allows us to rewrite the Coriolis
term $V=-{\bf{\omega\cdot\ell}}$ in the form $-2{\bf p\cdot A}$. It is
seen that the Hamiltonian (\ref{mfield}) lacks the
term $2M{\bf A}^2$ which is absolutely necessary for the gauge
invariance of Eqs. (\ref{gork}). However, since the two-body interaction
(\ref{int}) is invariant under the Gallileian transformation the conservation of
nucleon current is ensured.

The current density is expressed in terms of the Green's function $G$
as follows \cite{AGD}:
\begin{equation}
{\bf j}({\bf r})=\lim_{{\bf r'}\to{\bf r}}\oint\limits_C\frac{d\varepsilon}{2\pi i}
\left\{\frac{i\hbar}{2M}(\bigtriangledown_{\rm{\bf r}}-
\bigtriangledown_{\rm{\bf r'}}) - [\bf{\omega r}]\right\}
G({\bf r},{\bf r'},\varepsilon).
\label{cur}
\end{equation}
With this definition, we find
\begin{equation}
{\rm div}\ \!{\bf j}({\bf r})=\lim_{{\bf r'}\to{\bf r}}
\oint\limits_C\frac{d\varepsilon}{2\pi\hbar}[h_\omega({\bf r})-h_\omega^+({\bf r'})]
G({\bf r},{\bf r'},\varepsilon).
\end{equation}
Using the first of Eq. (\ref{gork}) and their complex conjugate equation
\begin{equation}
[\varepsilon-h_\omega^+({\bf r'})+\varepsilon_F]G({\bf r},{\bf r'},\varepsilon) +
\tilde\Delta^*({\bf r'})F({\bf r},{\bf r'},\varepsilon)=\delta({\bf r}-{\bf r'}),
\end{equation}
we finally obtain
\begin{equation}
{\rm div}\ \!{\bf j}({\bf r})=\oint\limits_C\frac{d\varepsilon}{2\pi\hbar}
[\tilde\Delta({\bf r})F^+({\bf r},{\bf r},\varepsilon)-
\tilde\Delta^*({\bf r})F({\bf r},{\bf r},\varepsilon)].
\label{curr}
\end{equation}
The right-hand side of this equality vanishes due to the third Gor'kov
equation. Because the latter is derived assuming
delta-interaction we should conclude that the form of a two-body
interaction is essential for obtaining the current conservation. In
particular, the commonly used monopole pairing interaction is not
invariant under the transformation (\ref{grad}). Therefore it does
not conserve the current density in a rotating nucleus. The case of
an arbitrary pairing interaction is considered in Ref.~\cite{Bel}.

\subsection{Matrix form of the Gor'kov equations}

The two-dimensional form of the Gor'kov equations is very convenient
to use in our calculations. Let us introduce the second pair of the
Green's functions $G^+({\bf r},{\bf r'},\varepsilon)$ and
$F({\bf r},{\bf r'},\varepsilon)$. It is easily proved \cite{GP} that
the four equations for these functions can be written in the matrix
form
\begin{equation}
\left(\begin{array}{cc}
\tilde\Delta^*& \varepsilon+h^+_\omega-\varepsilon_F \\
 \varepsilon-h_\omega+\varepsilon_F& \tilde\Delta \\
\end{array}\right)
\left(\begin{array}{cc}
F& G \\
G^+& F^+
\end{array}\right) = \hat 1\delta({\bf r}-{\bf r'}).
\label{matr}
\end{equation}
The operator $h_\omega$ involves the real $h$  and the imaginary $V$
parts. The former is the Hamiltonian of the deformed mean field,
whereas the latter denotes the cranking term,
$V=-\omega\ell_x$. Separating the quantity $\tilde\Delta$ into real
and imaginary parts,
\begin{equation}
\tilde\Delta=\Delta+\bar\Delta, \hspace{5mm}
\tilde\Delta^*=\Delta-\bar\Delta,
\end{equation}
we can rewrite Eq. (\ref{matr}) in the compact form:
\begin{equation}
(i\hat p +\Delta-\hat\sigma_1V-\hat\sigma_3\bar\Delta)
    \hat G({\bf r},{\bf r'},p) =\delta({\bf r}-{\bf r'}),
\label{matrx}
\end{equation}
where $\hat G$ is the matrix of the functions $G$ and $F$ involved
in Eq. (\ref{matr}),
$\hat p=\hat\sigma_1p+\hat\sigma_2(h-\varepsilon_F)$,
$\hat\sigma_\alpha$ are the Pauli matrices, and $p=-i\varepsilon$.
We omit the unit matrices before the terms with $\Delta$ and
$\delta({\bf r}-{\bf r'})$. The functions $G$ and $F$ can be written
as traces of $\hat G$ in the following way: \footnote{We use
the symbol $\rm tr$  in the space of single-particle states of the
Hamiltonian $h$, the symbol $\rm Sp$ in the spinor space, and
$\rm Tr$ in the combined space.}
$$
G({\bf r},{\bf r'},\varepsilon)=
\frac{1}{2}{\rm Sp}\{(\hat\sigma_1-i\hat\sigma_2)\hat G({\bf r},{\bf r'},p)\},
$$
\begin{equation}
F({\bf r},{\bf r'},\varepsilon)=
\frac{1}{2}{\rm Sp}\{(1+\hat\sigma_3)\hat G({\bf r},{\bf r'},p)\}.
\end{equation}
Therefore, the equation for $\tilde\Delta({\bf r})$ is
\begin{equation}
\tilde\Delta({\bf r})= g\int\limits_{C'}\frac{dp}{4\pi}
{\rm Sp}\{(1+\hat\sigma_3)\hat G({\bf r},{\bf r},p)\},
\label{gapeq}
\end{equation}
and the one-particle density matrix of the system is given by the
expression
\begin{equation}
\rho({\bf r})=\int\limits_{C'}\frac{dp}{4\pi}
{\rm Sp}\{(\hat\sigma_1-i\hat\sigma_2)\hat G({\bf r},{\bf r},p)\},
\label{dmatr}
\end{equation}
where the contour $C'$ is obtained from $C$ by the $90^\circ$
rotation.

\subsection{Perturbation theory}

We apply the Green's function method to calculate the parameter
$\cal B$. As follows from Eqs. (\ref{cranc}) and (\ref{cone}), this
requires the perturbation theory of third order in the cranking term
$V$. According to Ref. \cite{Mar}, a considerable computational effort
is needed to mold the result into a tractable form. The matrix
representation of the Gor'kov equations allows us to elaborate on
the elegant form of the perturbation theory which considerably
simplifies the calculations.

We now proceed to treat Eq. (\ref{matrx}) by the method of successive
approximation. Considering $V$ as a weak perturbation, we expand
the Green's function and the self-consistent quantities $\tilde\Delta$
and $\varepsilon_F$ in the powers of a small parameter:
$$
\hat G=\hat G_0+\hat G_1+\hat G_2+\hat G_3+ ...,
$$
\begin{equation}
\tilde\Delta=\Delta^{(0)}+\bar\Delta^{(1)}+\Delta^{(2)}
        +\bar\Delta^{(3)}+..., \hspace{5mm}
\varepsilon_F=\varepsilon_F^{(0)}+\varepsilon_F^{(2)}+...\ .
\end{equation}
The $n$th order corrections to the last two quantities are determined
by Eqs. (\ref{gapeq}) and (\ref{numb}), respectively. Since $V=-V^*$,
the corrections of odd order to $\tilde\Delta$ are purely imaginary
and those of even order are real. The effect of the second order
correction to $\varepsilon_F$ gives a negligible small contribution
in the second inertial parameter \cite{Mar}. Thus, we will use
the zero-order approximation for this quantity.

It is natural to work in the basis of the eigenfunctions
of the Hamiltonian (\ref{mfield}) without the cranking term,
\begin{equation}
(h-\varepsilon_F)\varphi_\nu({\bf r})=p_\nu\varphi_\nu({\bf r}),
\label{schr}
\end{equation}
where $p_\nu$ is the energy $\varepsilon_\nu$ of the single-particle
state $\nu$ relative to the Fermi energy, $p_\nu=\varepsilon_\nu-
\varepsilon_F$. In this basis, the $n$th correction to the Green's
function has the form
\begin{equation}
\hat G_n({\bf r},{\bf r'},p)=\sum_{\nu\nu'}\hat G^{(n)}_{\nu\nu'}(p)
\varphi_\nu({\bf r})\varphi^*_{\nu'}({\bf r'}).
\label{coord}
\end{equation}

First of all, we find the solution of the unperturbed equation (\ref{matrx}),
\begin{equation}
(i\hat p +\Delta)\hat G_0({\bf r},{\bf r'},p) =\delta({\bf r}-{\bf r'}),
\label{matr0}
\end{equation}
with the constant pairing gap $\Delta^{(0)}=\Delta$. Substituting
Eq. (\ref{coord}) into (\ref{matr0}), one finds
\begin{equation}
\hat G^{(0)}_{\nu\nu'}({\bf p_\nu})=
      -\frac{i\hat p_\nu-\Delta}{{\bf p}_\nu^2+\Delta^2}\delta_{\nu\nu'},
\label{0apr}
\end{equation}
where ${\bf p_\nu}(p,p_\nu)$ is the two-dimensional vector and
$\hat p_\nu=\hat\sigma_1p+\hat\sigma_2p_\nu$.
The gap equation (\ref{gapeq}) takes the simple form
\begin{equation}
1=g\sum_\nu\frac{1}{2E_\nu}|\varphi_\nu({\bf r})|^2,
\hspace{4mm} E_\nu=\sqrt{p^2_\nu+\Delta^2}.
\label{gap}
\end{equation}
The equation has the solution $\Delta=const$ for
the self-consistent potential with a flat bottom. In obtaining Eq.
(\ref{gap}) as well as in subsequent calculations, it is
essential to compute the traces of the
products of the Pauli matrices. We can readily see that the trace
of an odd number of matrices $\hat\sigma_1$ and $\hat\sigma_2$
vanishes and that of an even number is given by the expressions
\begin{equation}
\frac{1}{2}{\rm Sp}(\hat\sigma_\alpha\hat\sigma_\beta)=
\delta_{\alpha\beta},
\hspace{5mm}
\frac{1}{2}{\rm Sp}(\hat\sigma_\alpha\hat\sigma_\beta\hat\sigma_\gamma
\hat\sigma_\delta)=\delta_{\alpha\beta}\delta_{\gamma\delta}
 -\varepsilon_{\alpha\beta}\varepsilon_{\gamma\delta},
\hspace{2mm}...,
\end{equation}
where $\hat\varepsilon=i\hat\sigma_2$ is a fully antisymmetric matrix.

In the first order, Eq. (\ref{matrx}) involves the two perturbing terms
$V$ and $\bar\Delta^{(1)}$:
\begin{equation}
(i\hat p +\Delta)\hat G_1({\bf r},{\bf r'},p) =
[\hat\sigma_1V+\hat\sigma_3\bar\Delta^{(1)}]\hat G_0({\bf r},{\bf r'},p).
\end{equation}
The solution of this equation is obvious:
\begin{equation}
\hat G_1({\bf r},{\bf r'},p) = \int\hat G_0({\bf r},{\bf q},p)
 \hat W({\bf q})\hat G_0({\bf q},{\bf r'},p)d{\bf q},
\end{equation}
where the operator
\begin{equation}
\hat W = \hat\sigma_1V+\hat\sigma_3\bar\Delta^{(1)}
\end{equation}
is introduced to write the corrections to the unperturbed Green's
function in the simple symbolic form:
$$
\hat G_1=\hat G_0\hat W\hat G_0, \hspace{8mm}
\hat G_2=\hat G_0\hat W\hat G_0\hat W\hat G_0
    -\hat G_0\Delta^{(2)}\hat G_0,
$$
\begin{equation}
\hat G_3=\hat G_0\hat W\hat G_0\hat W\hat G_0
\hat W\hat G_0 - \hat G_0\hat W\hat G_0\Delta^{(2)}\hat G_0
-\hat G_0\Delta^{(2)}\hat G_0\hat W\hat G_0
+\hat G_0\hat\sigma_3\bar\Delta^{(3)}\hat G_0. \vspace{2mm}
\label{correct}
\end{equation} 
Here the integration over intermediate coordinates $\bf q$ is implied.
Using these formulas, one can prove by straightforward calculations
the following identities:
$$
\int\limits_{C'}\frac{dp}{2\pi}{\rm Sp}
\{\hat G_{2i+1}({\bf r},{\bf r'},p)\}=0,
\hspace{7mm}
\int\limits_{C'}\frac{dp}{2\pi}{\rm Sp}
\{\hat\sigma_1\hat G_{2i}({\bf r},{\bf r'},p)\}=0,
$$
\begin{equation}
\int\limits_{C'}\frac{dp}{2\pi}{\rm Sp}
\{\hat\sigma_2\hat G_{2i+1}({\bf r},{\bf r'},p)\}=0,
\hspace{7mm}
\int\limits_{C'}\frac{dp}{2\pi}{\rm Sp}
\{\hat\sigma_3\hat G_{2i}({\bf r},{\bf r'},p)\}=0.
\label{idens1}
\end{equation}

In order to find the self-consistent solution, we have to show how
$\tilde\Delta^{(n)}$ is obtained from $\hat G_n$. We will consider the
unperturbed pairing gap $\Delta$ as a parameter of the theory. This
allows us to eliminate the interaction constant $g$. Multiplying the
zero-order equation (\ref{gap}) by $\tilde\Delta^{(n)}({\bf r})$
we write the result in the symmetric form:
\begin{equation}
\Delta\tilde\Delta^{(n)}({\bf r})=g\int\limits_{C'}\frac{dp}{8\pi}
{\rm Sp}[\hat G_0({\bf r},{\bf r},p),\tilde\Delta^{(n)}({\bf r})]_+,
\end{equation}
where $[...]_+$ is the anticommutator of corresponding operators.
With this ansatz, the integral equation for even order corrections is
given by:
\begin{equation}
\int\limits_{C'}\frac{dp}{2\pi}{\rm Sp}
 \{2\Delta\hat G_{2i}({\bf r},{\bf r},p)
 - [\hat G_0({\bf r},{\bf r},p),\Delta^{(2i)}({\bf r})]_+\}=0,
\label{even}
\end{equation}
while that for odd order ones has the form:
\begin{equation}
\int\limits_{C'}\frac{dp}{2\pi}{\rm Sp}
\{2\Delta\hat\sigma_3\hat G_{2i+1}({\bf r},{\bf r},p)
 - [\hat G_0({\bf r},{\bf r},p),\bar\Delta^{(2i+1)}({\bf r})]_+\}=0.
\label{odd}
\end{equation}
Similarly, eliminating $g$ from the equations for $\bar\Delta^{(n)}$
and $\Delta^{(n-1)}$ ($n$ is odd) yields
\begin{equation}
\int\limits_{C'}\frac{dp}{2\pi}{\rm Sp}
\{\bar\Delta^{(n)}({\bf r})\hat G_{n-1}({\bf r},{\bf r},p)
 - \Delta^{(n-1)}({\bf r})\hat\sigma_3\hat G_n({\bf r},{\bf r},p)\}=0.
\label{idens2}
\end{equation}
In the same manner, with additional integration over $d{\bf r}$, we
can derive
\begin{equation}
\int\limits_{C'}\frac{dp}{2\pi}{\rm Tr}
\{\bar\Delta^{(n)}\hat\sigma_3\hat G_{n-2}(p)
 - \bar\Delta^{(n-2)}\hat\sigma_3\hat G_n(p)\}=0.
\label{idens3}
\end{equation}
Here $\rm Tr$ refers to the spinor and to the single-particle
spaces simultaneously.

The solution of the HFB equations presented in the form or the
successive approximations by the formulas (\ref{0apr}), (\ref{gap}),
(\ref{correct}), (\ref{even}), (\ref{odd}) takes into consideration
nonuniform pairing correlations induced by rotation. The nonuniform
pairing originates in higher orders of the perturbation theory, while the
nonrotating system is approximated by the constant pairing field. In
spite of an apparent eclecticism, our solution does not violate
the current density conservation, ${\rm div}\ \!{\bf j}=0$. This is obvious
in the zero-order approximation. In the even orders of the perturbation
theory ${\rm div}\ \!{\bf j}^{(2i)}$ vanishes due to the first and four
identities of (\ref{idens1}). The odd corrections to the quantity
(\ref{curr}) can be transformed into the expression
\begin{equation}
{\rm div}\ \!{\bf j}^{(2i+1)}({\bf r})=\sum_{l=0}^i\int\limits_{C'}
  \frac{dp}{2\pi}{\rm Sp}\{\bar\Delta^{(2l+1)}({\bf r})
   \hat G_{2i-2l}({\bf r},{\bf r},p)
  -\Delta^{(2l)}\hat\sigma_3\hat G_{2i-2l+1}({\bf r},{\bf r},p)\},
\end{equation}
which vanishes due to the equation (\ref{idens2}). Thus, we have
found the consistent solution, which
is certainly more general than initially supposed. It can be used for
an arbitrary single-particle potential without danger of coming to
contradiction (see also Ref. \cite{Mig}). We will apply this solution to
calculation of the second inertial parameter.

Finally, the following identities are useful in calculations:
\begin{equation}
2\Delta\int\limits_{C'}\frac{dp}{2\pi}{\rm Tr}\{\Delta^{(2i)}\hat G_{2i}(p)\}=
 \int\limits_{C'}\frac{dp}{2\pi}{\rm Tr}\{\Delta^{(2i)}[\Delta^{(2i)},
         \hat G_0(p)]_+\},
\label{idens4}
\end{equation}
\begin{equation}
2\Delta\int\limits_{C'}\frac{dp}{2\pi}{\rm Tr}\{\bar\Delta^{(2i+1)}
               \hat\sigma_3\hat G_{2i+1}(p)\}=
 \int\limits_{C'}\frac{dp}{2\pi}{\rm Tr}\{\bar\Delta^{(2i+1)}
               [\bar\Delta^{(2i+1)},\hat G_0(p)]_+\}.
\label{idens5}
\end{equation}
These identities are obtained by multiplying Eq. (\ref{even}) and (\ref{odd}) by
$\Delta^{(2i)}({\bf r})$ and $\bar\Delta^{(2i+1)}({\bf r})$,respectively, and
integrating over $d{\bf r}$.

\setcounter{equation}{0}
\section{Calculation of the second inertial parameter}
\label{sec3}
\subsection{General expression}
As follows from Eq. (\ref{cone}), the derivation of the parameter
${\cal B}$ is reduced to the calculation of $\beta$. This latter
quantity is convenient to deal with. The third-order correction to
the density matrix can be obtained from Eqs. (\ref{dmatr}) and
(\ref{correct}). If we take into account the third equation from
(\ref{idens1}), substitute
\begin{equation}
\hat\sigma_1\ell_x=-\frac{1}{\omega}(W-\hat\sigma_3\Delta^{(1)}),
\end{equation}
and use the identity (\ref{idens3}) with $n=3$, we get after some
simple algebraic calculations
\begin{equation}
\beta_s= \frac{1}{\omega^3}{\rm tr}\{\ell_x\rho^{(3)}\}=
 -\frac{1}{\omega^4}\int\limits_{C'}\frac{dp}{4\pi}{\rm Tr}\{\hat W
  \hat G_3(p)\} + \frac{1}{\omega^4}\int\limits_{C'}\frac{dp}{4\pi}
     {\rm Tr}\{\Delta^{(3)}\hat\sigma_3\hat G_1(p)\},
\label{fin0}
\end{equation}
where the subscript $s$ means that the relevant quantity refers to
the superfluid state. Here and below we will use $\Delta^{(2i+1)}$
($i=0,1$) instead of $\bar\Delta^{(2i+1)}$. Fortunately,
the terms with $\Delta^{(3)}$ are eliminated from (\ref{fin0}) after
inserting the expression for $\hat G_3$ (\ref{correct}). The resulting
formula involves the corrections $\Delta^{(1)}$ and $\Delta^{(2)}$
only. It is convenient to transform the terms with $\Delta^{(2)}$ into
a quadratic form in this quantity. Referring to the definition of the
function $\hat G_2$ from Eqs. (\ref{correct}), we find that the term
involving $\Delta^{(2)}$ becomes
$$
\int\limits_{C'}\frac{dp}{2\pi}{\rm Tr}\{\Delta^{(2)}\hat G_0(p)
\hat W\hat G_0(p)\hat W\hat G_0(p)\}
$$
\begin{equation}
=\int\limits_{C'}\frac{dp}{2\pi}{\rm Tr}\{\Delta^{(2)}\hat G_2(p)\}
  -\int\limits_{C'}\frac{dp}{2\pi}{\rm Tr}\{\Delta^{(2)}\hat G_0(p)
    \Delta^{(2)}\hat G_0(p)\}.
\label{fin1}
\end{equation}
With the help of the identity (\ref{idens4}) it is easy to show that
the first term on the right also transforms into a bilinear form of
$\Delta^{(2)}$. Combining this result with (\ref{fin0}) we obtain the
final expression for the parameter $\beta_s$:
$$
\beta_s=-\frac{1}{\omega^4}
\int\limits_{C'}\frac{dp}{4\pi}{\rm Tr}\{\hat W\hat G_0(p)
\hat W\hat G_0(p)\hat W\hat G_0(p)\hat W\hat G_0(p)\}
$$
\begin{equation}
 +\frac{1}{\omega^4\Delta}\int\limits_{C'}\frac{dp}{4\pi}{\rm Tr}\{2\Delta
      \Delta^{(2)}\hat G_0(p)\Delta^{(2)}\hat G_0(p) +
      \Delta^{(2)}[\hat G_0(p),\Delta^{(2)}]_+\}.
\label{fin}
\end{equation}
This is an exact formula for the calculation of the second inertial
parameter. The first term describes the joint effect of the Coriolis
force and the nonuniform pairing field $\Delta^{(1)}({\bf r})$ on
independent quasiparticle motion. In
the limit of the monopole pairing interaction, which corresponds to
the uniform pairing field ($\Delta^{(1)}=0$), this agrees with the
term ${\cal B}_c$ found by Marshalek \cite{Mar}. The second term
arises only due to the modification of the pairing. In the limit of
uniform pairing this term describes the Corriolis antipairing effect
(see Appendix A).

\subsection{Semiclassical approximation}

To proceed further we should find the first and second order
corrections to the pairing field. As shown in Appendix A, the solutions
of the corresponding integral equations are
\begin{equation}
\Delta^{(1)}({\bf r})= -\frac{i\hbar\omega}{2\Delta}D_1\dot\ell_x,
\hspace{7mm}
\Delta^{(2)}({\bf r})= \frac{\hbar^2\omega^2}{4\Delta^3}D_2\dot\ell^2_x,
\label{nuif}
\end{equation}
where $D_1$ and $D_2$ are the amplitudes, which are
found in a self-consistent way. In order to learn more about the
nonuniform pairing, we suppose that the self-consistent potential of a
deformed nucleus is of the form
\begin{equation}
U(\rho)=U\left(\frac{x^2+y^2}{a^2}+\frac{z^2}{b^2}\right),
\end{equation}
where $a$ and $b$ are the half-axes of a nuclear spheroid. Then we
obtain
\begin{equation}
\dot\ell_x=y\frac{\partial U}{\partial z}-z\frac{\partial U}{\partial y}=
 \frac{b^2-a^2}{a^2b^2}yzU'(\rho).
\end{equation}
Thus, the rotation induced pairing field is a function of the space
coordinates only. In the first order in rotation, the nonuniform
pairing field is proportional to the spherical harmonic $Y_{2\pm 1}$.
This was a motive for introducing the quadrupole pairing (see
\cite{Muh} and references therein). The second correction
$\Delta^{(2)}$ shows, however, that higher multipoles are also
involved in nonuniform pairing.

Using the expression for $\Delta^{(1)}$ and the obvious formula for
matrix elements 
$\hbar\dot\ell^x_{\nu\nu'}=i(p_\nu-p_{\nu'})\ell^x_{\nu\nu'}$, we can
represent each of 16 sums in the first term of Eq. (\ref{fin}) in the
general form
\begin{equation}
\sum\!\ell^x_{12}\ell^x_{23}\ell^x_{34}\ell^x_{41}
\int\limits_{C'}\frac{dp}{2\pi}\frac{{\cal Q}_4(p;p_1,p_2,p_3,p_4)}
{({\bf p}_1^2+\Delta^2)({\bf p}_2^2+\Delta^2)({\bf p}_3^2+\Delta^2)
({\bf p}_4^2+\Delta^2)},
\label{sum}
\end{equation}
where the summation indices 1,2,3,4 refer to the single-particle
states $\nu$ with the energy $p_\nu$ of the Schr\"odinger equation
(\ref{schr}) and ${\cal Q}_4$ is a polynomial of the fourth power in
$(p,p_\nu)$ and $D_1$, which is derived from calculating the trace of
the product of the Pauli matrixes and factors $(i\hat p_\nu-\Delta)$.

To evaluate this sum we use the method proposed by Migdal \cite{Mig}.
Let us note first that at the fixed state 1 the indexes 2, 3, and 4
take only few values permitted by the selection rules for the matrix
elements of the operator $\ell_x$. After integration over $dp$ we
obtain the function of the variables $p_1, ..., p_4$. When
considered as the function of the variable $p_1$, it has for fixed
differences $p_1-p_2$, $p_1-p_3$, and $p_1-p_4$ a sharp
maximum at the Fermi surface with an approximate width $\Delta$.
Since the average level spacing for ND nuclei,
$\delta\varepsilon\sim \varepsilon_F/A$, is very small compared
to $\Delta\sim\varepsilon_F/A^{2/3}$, a large number
($\sim A^{1/3}$) of levels fall within the interval $\Delta$. For 
this reason we can make with semiclassical accuracy the following
substitution in the sum (\ref{sum}):
$$
\int\limits_{C'}\frac{dp}{2\pi}\frac{
{\cal Q}_4(p;p_1,p_2,p_3,p_4)}{({\bf p}_1^2+\Delta^2)
({\bf p}_2^2+\Delta^2)({\bf p}_3^2+\Delta^2)({\bf p}_4^2+\Delta^2)}
$$
\begin{equation}
\to \delta(\varepsilon_1-\varepsilon_F)\int\frac{d{\bf p}_1}{2\pi}
\frac{{\cal Q}_4( p;p_1,p_2,p_3,p_4)}
{({\bf p}_1^2+\Delta^2)({\bf p}_2^2+\Delta^2)({\bf p}_3^2+\Delta^2)
({\bf p}_4^2+\Delta^2)}.
\label{subst}
\end{equation}
Similarly, the second term of Eq. (\ref{fin}) can be approximated by
the expression
\begin{equation}
\frac{1}{\omega^4}\sum|\Delta^{(2)}_{12}|^2
\int\frac{d{\bf p}_1}{4\pi}\frac{(p_1-p_2)^2+4\Delta^2}
  {({\bf p}_1^2+\Delta^2)({\bf p}_2^2+\Delta^2)}
   \delta(\varepsilon_1-\varepsilon_F).
\label{sum1}
\end{equation}

When calculating these integrals over $d{\bf p}_1=dpdp_1$,
it is convenient to use the Feynman covariant integration method
\cite{Fyn} because good convergence of the integrals
allows the integration over $dp_1$ to be extended from $-\infty$ to $\infty$.
For details of calculations see Appendix B. The integral in
(\ref{subst}) depends on the three independent differences $p_1-p_2$,
$p_1-p_3$, and $p_1-p_4$. To represent the final result of the
semiclassical approximation in a symmetrical form, we introduce the
six energy differences
$p_{\nu\nu'}=p_\nu-p_{\nu'}$, $\nu<\nu'$.
Collecting all the integrals of the first and the second terms in
(\ref{fin}), we find
\begin{equation}
\beta_s\!=\frac{1}{4\Delta^2}\sum\!\ell^x_{12}
  \ell^x_{23}\ell^x_{34}\ell^x_{41}
    F(x_{12},x_{23},x_{34},x_{41};x_{13},x_{24})
     \delta(\varepsilon_1\!\!-\!\varepsilon_F\!),
\label{spur}
\end{equation}
where $\delta$ function denotes that summation over the state 1 is
substituted, according to the semiclassical approximation, by
integration over its quantum numbers.

The function $F$, which depends on the six dimensionless differences
$x_{\nu\nu'}=(\varepsilon_\nu\!-\!\varepsilon_{\nu'})/2\Delta$, is
divided into two parts:
\begin{equation}
F\!=\!f(x_{12},x_{23},x_{34},x_{41};x_{13},x_{24})
      +8D^2_2 x_{12}x_{23}x_{34}x_{41}h(x_{13}).
\label{fun}
\end{equation}
The first one is  relevant for the first term of Eq. (\ref{fin}). It
is convenient to represent this function in the form
\begin{equation}
f\!=-(1+\hat P_1+\hat P_2+\hat P_3)G+\!(1+\hat P_1)H,
\label{fun1}
\end{equation}
where
$$
G\!=\!\frac{g(x_{12})}{x_{13}x_{23}x_{24}x_{41}}
   \!\left\{\!(1\!-\!D_1 x^2_{12})\! \left[1\!\!+\!x^2_{12}\!+
      \!x_{23}x_{41}\!-\!D_1[x^2_{23}(1\!-\!x_{12}x_{24})\right.\right.
$$
$$
+\left.\left. x^2_{41}(1\!+\!x_{12}x_{13})]-
    \!D^2_1x_{23}x_{34}x_{41}(x_{23}\!+\!x_{41})
     -D^3_1x_{12}x^2_{23}x_{34}x^2_{41}\right]\right.
$$
\begin{equation}
-\left.D_1(x_{34}-D_1x_{12}x_{23}x_{41})(x_{34}
    +x_{12}x_{13}x_{24}-D_1x_{12}x_{23}x_{41})\right\}
\end{equation}
and
\begin{equation}
\hspace{3pt}
H\!=\!\frac{h(x_{13})}{x_{12}x_{23}x_{34}x_{41}}\!\left[1\!-
\!D_1( x^2_{12}\!+\!x^2_{23}\!+\!x^2_{34}\!+\!x^2_{41})
+D^2_1( x_{12}x_{41}\!+\!x_{23}x_{34})^2\right].
\end{equation}
Here the functions
\begin{equation}
g(x)=\frac{{\rm argsh}x}{x\sqrt{1+x^2}},\hspace{7mm}h(x)= (1+x^2)g(x)
\label{fuhg}
\end{equation}
are associated with the Migdal moment of inertia \cite{Mig}.
The amplitudes $D_1$ and $D_2$ of the nonuniform pairing field are
determined by Eqs. (\ref{amp1}) and (\ref{amp2}), respectively.
The operator $\hat P_i$ permutes the indices $\nu$ of single-particle
states in all the dimensionless differences, on which the functions
$G$ and $H$ depend. When applied to $x_{\nu\nu'}$, we get
\begin{equation}
\hat P_ix_{\nu,\nu'}=x_{\nu+i,\nu'+i},
\end{equation}
subject to $\nu\ \!{\rm mod}\ \!4=\nu$. It is easy to prove the
following symmetry properties of the function $F$:
$$
\hat P_1F=\hat P_2F=\hat P_3F=F,
$$
\begin{equation}
F(x_{12},x_{41},x_{34},x_{23};-x_{24},-x_{13})=
F(x_{12},x_{23},x_{34},x_{41};x_{13},x_{24}).
\label{sym}
\end{equation}

The above formulas give the semiclassical expression
for the second inertial parameter
in the superfluid phase. The solution takes into
account the effect of rotation on the Cooper pairs in the
gauge-invariant form. The result is expressed entirely in terms
of matrix elements and corresponding energy differences providing the
constant pairing gap $\Delta$ is fixed for a nonrotating nucleus. It is
valid for an arbitrary nuclear mean field with a stable
deformation. This allows one to study an interplay between
rotation, pairing correlations, and mean field deformation in ND and
SD bands.

We first estimate the quantity $\beta_s$ and find the small parameter
of the perturbation theory we used. To get an
estimate of $\beta_s$ for ND bands, we observe that the matrix
element $\ell^x_{\nu\nu'}$ has the two types of transitions $\nu\to\nu'$:

(i) transitions inside the $\cal N$-shell (close transitions), for which
energy differences are $p_{\nu\nu'}=d_1\sim\varepsilon_FA^{-2/3}$,
and the maximal value $L$ of the matrix element $\ell^x_{\nu\nu'}$
is related to a transition between states of a $j$-shell;

(ii) transitions between shells with major quantum numbers
$\cal N$ and ${\cal N}\pm 2$ (distant transitions) with
$p_{\nu\nu'}=d_2\sim\varepsilon_FA^{-1/3}$ and
$\ell^x_{\nu\nu'}\sim LA^{-1/3}$. \\
For the state 1, there are the three groups of terms in the sum
(\ref{spur}), which are classified according to different combinations
of the close and distant transitions in the product of the four matrix
elements $\ell^x_{\nu\nu'}$. Those involving four close transitions
have all the dimensionless differences $x_{\nu\nu'}\sim 1$ and
consequently $F\sim 1$.\footnote{The necessary
estimation for the amplitudes of the uniform pairing field,
$D_1\sim D_2\sim [\ln{(d_2/\Delta)}]^{-1}\sim 1$, can be obtained from
Eqs. (\ref{amp1}) and (\ref{amp2}).} Thus, the
contribution of these terms to the sum (\ref{spur}) is of the order
$L^4$. For terms with four distant transitions
($x_{\nu\nu'}\sim A^{1/3}$), we have $F\sim A^{4/3}$. However, this
large factor is compensated by the product of small matrix elements
$\ell^x_{\nu\nu'}$. The same compensation takes place in the remaining
terms with two close and two distant transitions, for which
$F\sim A^{2/3}$. Therefore, the contributions of all terms in the sum
(\ref{spur}) are of the same order of magnitude $L^4$, and we can
make the following estimation:
\begin{equation}
\beta_s\sim\frac{1}{4\Delta^2}
 \sum\!\ell^x_{12}\ell^x_{23}\ell^x_{34}\ell^x_{41}
    \delta(\varepsilon_1\!\!-\!\varepsilon_F\!)=
\frac{1}{4\Delta^2}
\sum_1(\ell^4_x)_{11}\delta(\varepsilon_1\!\!-\!\varepsilon_F\!).
\label{est1}
\end{equation}
Calculation of the last sum within the Tomas-Fermi approximation gives
\begin{equation}
\beta_s\sim \frac{3M}{20\Delta^2}
\int n({\bf r})p^2_F({\bf r})(y^2+z^2)^2d{\bf r}.
\label{est2}
\end{equation}
In this calculation we used the procedure described in Ref. \cite{Mig},
which includes averaging over the direction of the nucleon impulse
and an utilization of the ansatz
\begin{equation}
\sum_1\varphi^*_1({\bf r})\varphi_1({\bf r})
  \delta(\varepsilon_1\!\!-\!\varepsilon_F\!)
     =\frac{3M}{p^2_F({\bf r})}n({\bf r}),
\label{ans}
\end{equation}
where $n({\bf r})=Cp^3_F({\bf r})$ $(C=const)$ is the nucleon density
and $p_F({\bf r})=\sqrt{2M[\varepsilon_F-U({\bf r})]}$. Comparing
(\ref{est2}) with the rigid-body moment of inertia
\begin{equation}
\Im_{\rm rig}=\int n({\bf r})(y^2+z^2)d{\bf r},
\label{rig}
\end{equation}
we obtain $\beta_s\sim\Im_{\rm rig}(p_FR/\Delta)^2\sim\Im_{\rm rig}
(\hbar j_F/\Delta)^2$, where $R$ is the mean square radius of a nucleus,
$p_F$ and $j_F\sim A^{1/3}$ are the mean impulse and the mean angular
momentum of a nucleon on the Fermi surface. Thus, the parameter
$\beta_s$ has the order of magnitude $\hbar^4A^{11/3}/\varepsilon^3_F$.
With these results we can get from (\ref{Hexp}) the perturbation
parameter, $\beta_s\omega^2/\alpha\sim(\hbar\omega j_F/\Delta)^2$. An
application of the perturbation theory implies that this value is small,
i. e., the Coriolis interaction is smaller than a
two-quasiparticle excitation energy. One can say that the perturbation
theory is valid for adiabatic rotation. It is clear from Eq.
(\ref{cone}) that ${\cal B}_s\sim\varepsilon_FA^{-3}$ and
${\cal B}_s/{\cal A}\sim A^{-4/3}$. The above estimations refer to
ND nuclei in the ground state where pairing correlations are
reasonably strong, $\Delta\sim\varepsilon_FA^{-2/3}$.

\setcounter{equation}{0}
\section{The model of anisotropic oscillator potential}
\label{sec4}

In order to obtain quantitative results, we model the real
self-consistent nuclear field as the axially deformed  oscillator
potential with the frequencies $\omega_z$ along the symmetry axis
and $\omega_x$ in the perpendicular plane:
\begin{equation}
U_{\rm osc}({\bf r})=\frac{M}{2}[\omega^2_x(x^2+y^2)+\omega^2_zz^2].
\label{oscp}
\end{equation}
The use of this simplified
model is justified by a possibility of deriving an exact analytical
expression for the parameter $\beta_s$. It is known also that the
model reproduces the experimental values of the rotational constants
${\cal A}$ and ${\cal B}$ for ND bands. Therefore we can expect the
model to provide an useful insight in the rotational regime
at superdeformation.

In an anisotropic oscillator potential the matrix element $\ell^x_{\nu\nu'}$ is
non-zero for four transitions. The two close transitions have the
energy differences $d_1=\pm\hbar(\omega_x-\omega_z)$, whereas the
distant ones have $d_2=\pm\hbar(\omega_x+\omega_z)$. The
corresponding dimensionless parameters are
\begin{equation}
\nu_{1,2}\!=\!\frac{\hbar(\omega_x\!\mp\!\omega_z)}{2\Delta}\!=
\!\frac{k\mp1}{2\xi k^{2/3}},\hspace{5mm}
 \xi\!=\!\frac{\Delta}{\hbar\omega_0},
\label{osc}
\end{equation}
where $\hbar\omega_0\!=\!41A^{-1/3}$\,MeV. Hereafter we use
the axis or frequency ratio $k=b/a=\omega_x/\omega_z$ and the volume
conservation condition $\omega^2_x\omega_z=\omega^3_0$. It is also
convenient to substitute the operator $\ell_x$ with its time derivative
\begin{equation}
\dot\ell_x=M(\omega^2_x-\omega^2_z)yz,
\end{equation}
which has the same selection rules.

For the fixed state 1, the sum (\ref{spur}) involves 6 terms with the
four close transitions, 6 terms with the four distant transitions,
and 24 terms with the two close and two distant transitions. The
products of four matrix elements are equal with the semiclassical
accuracy for all 36 terms of the sum,
$\dot\ell^x_{12}\dot\ell^x_{23}\dot\ell^x_{34}\dot\ell^x_{41}
\approx\frac{1}{36}(\dot\ell^4_x)_{11}$. Summation of all these terms
gives
\begin{equation}
\beta_s = \frac{\hbar^4\Phi_n(\nu_1,\nu_2)}{18(2\Delta)^6\nu_1^4}
\sum_1(\dot\ell^4_x)_{11}\delta(\varepsilon_1\!\!-\!\varepsilon_F\!).
\label{ndbs1}
\end{equation}
The function $\Phi_n$ is the sum of the functions $F$ corresponding
to all 36 combinations of the close and distant transitions. The sum
can be simplified by using the symmetry properties (\ref{sym}). It is
convenient to represent the resulting function in the form
\begin{equation}
\Phi_n(\nu_1,\nu_2)={\cal F}(\nu_1,\nu_2)+2D^2_2{\cal H}(\nu_1,\nu_2),
\label{sdbs}
\end{equation}
where
$$
{\cal F}(\nu_1,\nu_2)=f(\nu_1,-\nu_1,\nu_1,-\nu_1;0,0)
 +2f(\nu_1,\nu_1,-\nu_1,-\nu_1;2\nu_1,0)
$$
$$
+2(\nu_1/\nu_2)^2[2f(\nu_1,\nu_2,-\nu_2,-\nu_1;\nu_1+\nu_2,0)
 +2f(\nu_1,-\nu_2,\nu_2,-\nu_1;\nu_1-\nu_2,0)
$$
$$
  +f(\nu_1,\nu_2,-\nu_1,-\nu_2;\nu_1+\nu_2,-\nu_1+\nu_2)
    +f(\nu_1,-\nu_2,-\nu_1,\nu_2;\nu_1-\nu_2,-\nu_1-\nu_2)]
$$
\begin{equation}
+(\nu_1/\nu_2)^4[f(\nu_2,-\nu_2,\nu_2,-\nu_2;0,0)
 +2f(\nu_2,\nu_2,-\nu_2,-\nu_2;2\nu_2,0)],
\label{nsf}
\end{equation}
and the amplitudes $D_1$, $D_2$, and the functions ${\cal H}$ are
determined by Eqs. (\ref{amh}), (\ref{amh1}), and (\ref{amh2}),
respectively. The two terms in (\ref{sdbs}) describe the two distinct
effects of the Coriolis force: the rotation-quasiparticle interaction
and the modification of pairing.

In the Tomas-Fermi approximation, we have according to the ansatz
(\ref{ans})
\begin{equation}
\sum_1(\dot\ell^4_x)_{11}\delta(\varepsilon_1\!\!-\!\varepsilon_F\!)
    =3CM\int\dot\ell^4_x({\bf r})
       \sqrt{2M[\varepsilon_F-U_{\rm osc}({\bf r})]}d{\bf r}.
\label{sum3}
\end{equation}
Combining the result of integration with the expressions for the
rigid-body moment of inertia (\ref{rig}) and the mean level density
near the Fermi surface
\begin{equation}
\rho_F=\int d{\bf r}\sum_1\varphi^*_1({\bf r})\varphi_1({\bf r})
  \delta(\varepsilon_1\!\!-\!\varepsilon_F\!)
     =3CM\int\sqrt{2M[\varepsilon_F-U_{\rm osc}({\bf r})]}d{\bf r},
\label{rho}
\end{equation}
we obtain
\begin{equation}
\sum_1(\dot\ell^4_x)_{11}\delta(\varepsilon_1\!\!-\!\varepsilon_F\!)
  =\frac{18(\omega_x-\omega_z)^4(1+k)^4}
          {5\rho_F(1+k^2)^2}\Im^2_{\rm rig}.
\label{sum2}
\end{equation}

In the case of a normal deformation, the quantity $k$ is close to unity
and consequently $\nu_2\gg\nu_1$. Inserting (\ref{sum2}) into
(\ref{ndbs1}) and using the above approximations we get for ND bands
\begin{equation}
\beta_s({\rm ND}) = \frac{\Im_{\rm rig}^2\Phi_{nc}(\nu_1,\nu_2)}
   {5\rho_F\Delta^2}.
\label{ndbs}
\end{equation}
The function $\Phi_{nc}$ approximates $\Phi_n$ in "the close
transition limit":
\begin{equation}
\Phi_{nc}(\nu_1,\nu_2)={\cal F}_c(\nu_1,\nu_2)
      +2D^2_2{\cal H}_c(\nu_2),
\label{ndf} \vspace{-5mm}
\end{equation}
where
\begin{equation}
{\cal F}_c(\nu_1,\nu_2)\!=\!f(\nu_1,-\nu_1,\nu_1,-\nu_1;0,0)
   +2f(\nu_1,\nu_1,-\nu_1,-\nu_1;2\nu_1,0).
\label{fcl}
\end{equation}
$\Phi_n$ depends on $\nu_2$ only through the amplitudes $D_1$, $D_2$, 
and the function
\begin{equation}
{\cal H}_c(\nu_2)=8+8\ln2\nu_2+\ln4\nu_2.
\end{equation}
The level density near the Fermi surface can be obtained by combining
(\ref{rho}) with the number of nucleons $A=\int n({\bf r})d{\bf r}$.
The result, $\rho_F=3A/\varepsilon_F$, depends on the Fermi energy
which is found from the volume conservation condition $a^2b=R^3$.
The expression (\ref{ndbs}) has been obtained in Ref. \cite{GP}.

Figure 2a shows that $\Phi_{nc}$ approximates the exact
function $\Phi_n$ very well. It shows also that the contribution of
the rotation-quasiparticle interaction is small compared to that of
the pairing modification. This result is explained by the
interference of the two effects: the Coriolis force and the
nonuniform pairing field $\Delta^{(1)}$. Neglecting the latter
results in compatible contributions of the two terms of
Eq. (\ref{ndf}) as it is seen from  Fig. 2b. This result is
consistent with the Marshalek calculations \cite{Mar}.

For superdeformed nuclei the parameters $\nu_1$ and $\nu_2$ are both
large. Thus, we should expect decrease of $\beta_s$. It is
convenient to rewrite Eq.\ (\ref{ndbs1}) by introducing, according
to (\ref{osc}), the new parameters $\xi$ and $k$ instead of $\nu_1$
and $\nu_2$. We express the sum (\ref{sum3}) in terms of
the rigid-body  moment of inertia and the number of nucleons $A$:
\begin{equation}
\sum_1(\dot\ell^4_x)_{11}\delta(\varepsilon_1\!\!-\!\varepsilon_F\!)
  =\frac{24(\omega_x-\omega_z)^4k^{2/3}(1+k)^4}
          {5(1+k^2)^3A^2}\omega^2_0\Im^3_{\rm rig}.
\label{sum4}
\end{equation}
We have now:
\begin{equation}
\beta_s({\rm SD}) = \frac{k^{2/3}(1+k)^4}{15\hbar^2(1+k^2)^3A^2}
\Im^3_{\rm rig}\Phi(\xi,k),
\hspace{7mm} \xi^{2}\Phi(\xi,k)=\Phi_n(\nu_1,\nu_2).
\label{spur1}
\end{equation}
The function $\Phi$, along with its limiting case of the uniform
pairing, is shown in Fig. 3. It can be seen that nonuniform pairing
reduces $\beta_s({\rm SD})$ even more than $\beta_s({\rm ND})$.

Taking $k=2$ and $\Delta=0.5$\,MeV ($\xi=0.065$) as the
representative parameters for SD bands, we find from Fig. 3 that
$\Phi\sim 1$. This yields the following estimation:
$\beta_s({\rm SD})\sim\hbar^4(A/\varepsilon_F)^3\sim
\beta_s({\rm ND})A^{-2/3}$ and ${\cal B}/{\cal A}\sim A^{-2}$. The
last estimation is correct for the SD bands in the $A\sim$190 mass
region where ${\cal B}/{\cal A}\sim 10^{-5}$. Yet  it overestimates
the experimental value for $^{84}$Zr(1) ($\sim10^{-5}$) and
$^{144}$Gd(1), $^{152}$Dy(1) ($\sim 10^{-6}$). The later two bands
have the smallest value of this ratio among all SD mass regions.
Thus, a high deformation and nonuniform pairing do not solve the
problem of the SD band regularity.

\subsection{Limiting cases}

The limiting cases most interesting to us are: strong pairing,
uncorrelated nucleons, and extremely large deformations. Reference 
\cite{GP} considers the limit of small deformations.

For a very strong pairing ($\Delta\gg\hbar\omega_0$), the size of
the Cooper pair $R\hbar\omega_0/\Delta$ becomes much less than
the nuclear radius $R$. Rotation of such nucleus  is described by the
hydrodynamic equations of the ideal liquid \cite{B/M}, according to
which the second inertial parameter vanishes. In the following analysis,
the key aspect is the nonuniform pairing.
For strong pairing, the quantities $\Delta^{(1)}$ and $\Delta^{(2)}$
are proportional to $\Delta$ because, as follows from Eqs. (\ref{amp1})
and (\ref{amp2}), $D_1\sim\Delta^2$ and $D_2\sim \Delta^4$. Therefore
this limit is instructive since it allows to check the solution of the
integral equation for $\Delta^{(2)}$.\footnote{The solution
for $\Delta^{(1)}$ has been verified in Ref. \cite{Mig}
by obtaining the hydrodynamic moment of inertia.}

In the strong pairing limit the parameters $\nu_1$ and $\nu_2$ are
small. It is possible to simplify the function $F$ by expanding
$g(x_{\nu\nu'})$ and $h(x_{\nu\nu'})$ in power series of
$x_{\nu\nu'}$, and then approximate with a necessary accuracy by
\begin{equation}
F={\cal P}_2+D_1{\cal P}_4+D^2_1{\cal P}_4+D^3_1{\cal P}_6+
D^4_1{\cal P}_8,
\end{equation}
where ${\cal P}_n$ is the polynomial of the $n$th power in
$x_{\nu\nu'}$. With this function, performing a calculation similar to
the one we used to obtain $\Phi_n$, we find the limiting value
\begin{equation}
\lim_{\Delta\to\infty}\Phi_n(\nu_1,\nu_2)=
-\frac{64}{3}\left(\frac{\omega_x-\omega_z}{\omega_x+\omega_z}\right)^2.
\end{equation}
Combining this result with Eqs. (\ref{ndbs1}) and (\ref{sum4}) gives
\begin{equation}
\beta_s\sim-\frac{\Im^3_{\rm rig}}{(\hbar A)^2}
 \left(\frac{b^2-a^2}{b^2+a^2}\right)^2
       \left(\frac{\hbar\omega_0}{\Delta}\right)^2.
\end{equation}
Thus, the parameter $\beta_s$ vanishes in the hydrodynamic limit.

The rotation of a very elongated nucleus with $k=b/a\gg 1$ exhibits
some interesting physics. For this limit, the parameters $\nu_1$ and
$\nu_2$ are approximately equal
\begin{equation}
\nu_1=\nu\pm\delta\nu, \hspace{5mm} \nu=\frac{\omega_x}{2\Delta},
  \hspace{5mm} \delta\nu=\frac{\omega_z}{2\Delta},
\end{equation}
where $\nu\gg 1$ and $\delta\nu/\nu=a/b\ll 1$. The nonuniform
pairing is also important in this case because the small amplitude
$D_1\approx1/\nu^2$ is compensated by the large value of $\nu^2$.
As a result, the kinematic moment of inertia, which is the sum of the
standard cranking-model term and the Migdal one, is close to the
rigid body value:
\begin{equation}
\Im^{(1)}=\Im_{\rm rig}\left[1-\left(\frac{a}{b}\right)^2
  \frac{10}{\nu^2}\ln{2\nu}\right].
\end{equation}
For the first approximation $\nu_1=\nu_2$, the function (\ref{sdbs})
vanishes, $\Phi_n(\nu,\nu)=0$. The next term of its expansion in
$\delta\nu$ gives the estimation of the second inertial parameter
$\beta_s\sim(a/b)^2$. We can say that a strongly elongated nucleus
in the superfluid phase has the rotational regime which is close to
the rigid-body rotation. The deviation are of the order $(a/b)^2$.
The physical interpretation of this phenomenon is straightforward:
all nucleons of a needle shaped nucleus with exclusion of the small
sphere of the radius $a$ in its center are involved in
rotational motion.

Let us now consider the normal phase. The right-hand side of
Eq.\,(\ref{spur}) vanishes in the limiting case $\Delta=0$. This
result is an artifact of the semiclassical approximation used in
deriving the expression (\ref{spur}). The correct formula obtained
from Eq. (\ref{fin}) with the limiting values of the Bogolubov
amplitudes ($u_\nu=0$, $v_\nu=1$ for $\rho_\nu=1$ and $u_\nu=1$,
$v_\nu=0$ for $\rho_\nu=0$, where $\rho_\nu$ is the nucleon occupation
numbers) is
\begin{equation}
\beta^{({\rm sp})}_n =- \sum\ell^x_{12}\ell^x_{23}
    \ell^x_{34}\ell^x_{41}\sum^3_{i=0}\!\hat P_i\left\{\frac{\rho_1}
{(\varepsilon_1-\varepsilon_2)(\varepsilon_1-\varepsilon_3)
    (\varepsilon_1-\varepsilon_4)}\right\}.
\label{spur2}
\end{equation}
This expression describes the effect of the Coriolis force on
single-particle motion. It will be shown in the next section that
the cancellation of the leading terms in the sum of (\ref{spur2})
substantially reduces this quantity compared to $\beta_s$.

\subsection{The second inertial parameter for uncorrelated nucleons}

In this subsection we estimate the parameter $\beta$ in the normal
phase. In view of the cancelation mentioned above, we have to take
into account the centrifugal stretching effect, which happens to be
the same order of magnitude as (\ref{spur2}). As with the superfluid
phase, we will use the Green's function technique.\footnote{There is
an alternative method based on minimization of the
system energy in the rotating frame as a function of the oscillator
frequencies $\omega_x, \omega_y, \omega_z$, and the rotational
frequency $\omega$ under the constraint of constant volume. For
the fixed occupation of single particle states, this method gives the
same result as the one obtained below.} Our starting point is the
equations in the Hartree approximation\footnote{This approximation
is adequate for a separable two-body interaction we will use.}
\begin{equation}
[\varepsilon-h_\omega({\bf r})-{\cal V}({\bf r})]
   G({\bf r},{\bf r'},\varepsilon)=\delta({\bf r}-{\bf r'}),
\end{equation}
where $h_\omega$ is the cranked single-particle Hamiltonian
(\ref{mfield}) with the oscillator potential (\ref{oscp}) and
\begin{equation}
{\cal V}({\bf r})=\int d{\bf q}v_q({\bf r},{\bf q})
\oint\limits_C\frac{d\varepsilon}{2\pi i}G({\bf q},{\bf q},\varepsilon)
\end{equation}
is the self-consistent potential. We assume that the two-body
residual interaction $v_q$ is the effective quadrupole one,
\begin{equation}
v_q({\bf r},{\bf r}_1)=-\frac{\chi}{2}\sum_\mu(-1)^\mu q_{2\mu}({\bf r}'')
      q_{2-\mu}({\bf r}''_1),
\end{equation}
where the quadrupole moment $q_{2\mu}$ is defined in terms of the doubly 
stretched coordinates
\begin{equation}
{\bf r}''={\bf i}\frac{\omega_x}{\omega_0}x+
{\bf j}\frac{\omega_y}{\omega_0}y+{\bf k}\frac{\omega_z}{\omega_0}z.
\end{equation}
The interaction strength $\chi$ is determined in a self-consistent way
as follows:
\begin{equation}
\chi=\frac{4\pi\omega_0^2M}{5{\rm tr}\{({\bf r}'')^2\rho\}}.
\label{str}
\end{equation}
This interaction provides the full self-consistency for deformed nuclei
\cite{SK}.

As usual, we proceed to treat the cranking term $V$ with the
perturbation theory by expanding the Green's function and the
self-consistent potential in the series
\begin{equation}
G=G_0+G_1+G_2+G_3+..., \hspace{7mm}  {\cal V}=
{\cal V}^{(0)}+{\cal V}^{(1)}+{\cal V}^{(2)}+{\cal V}^{(3)}+...\ .
\end{equation}
The unperturbed Green's function is
\begin{equation}
G_0({\bf r},{\bf r}_1,\varepsilon)=\sum_\nu G_{\nu}(\varepsilon)
\varphi_\nu({\bf r})\varphi^*_{\nu}({\bf r}_1), \hspace{5mm}
G_\nu(\varepsilon) = \frac{1}{\varepsilon - \varepsilon_\nu +
     i\delta(1-2\rho_\nu)},
\end{equation}
with $\delta\to +0$. The occupation numbers $\rho_\nu$ refer to a
nonrotating nucleus. We may notice that under the self-consistent
condition
\begin{equation}
\omega_x\Sigma_x=\omega_x\Sigma_y=\omega_z\Sigma_z,
\hspace{7mm} \Sigma_{x,y,z}=\sum_\nu(n_{x,y,z}+1/2)_\nu\rho_\nu
\end{equation}
($n_x, n_y, n_z$ are the oscillator quantum numbers)
${\cal V}^{(0)}=0$. Thus, the average potential is modified
by rotation only.

The odd corrections to the self-consistent potential
are lacking, ${\cal V}^{(2i+1)}=0$, due to the different symmetry
properties of the operators $q_{2\mu}$ and $\ell_x$ under the time
reversal. Consequently the third order correction to the Green's
function is expressed as
\begin{equation}
G_3=G_0VG_0VG_0VG_0+G_0VG_0{\cal V}^{(2)}G_0+
       G_0{\cal V}^{(2)}G_0VG_0.
\end{equation}
The first term generates the interaction of rotation with 
single-particle motion. It yields the quantity 
$\beta^{({\rm sp})}_n$ (\ref{spur2}). The last two
are responsible for the centrifugal-stretching effect which is
described by the expression
\begin{equation}
\beta^{({\rm str})}_n=-\frac{2}{\omega^4}
 \oint\limits_C\frac{d\varepsilon}{2\pi i}
{\rm tr}\{{\cal V}^{(2)}G_0(\varepsilon)VG_0(\varepsilon)
     VG_0(\varepsilon)\},
\label{ctf}
\end{equation}
where the correction to the mean potential ${\cal V}^{(2)}$ is
obtained from the equation
\begin{equation}
{\cal V}^{(2)}({\bf r})\!\!=\!\!-\chi\!\sum_\mu (-1)^\mu q_{2-\mu}({\bf r})\!\!
\oint\limits_C\!\frac{d\varepsilon}{2\pi i}
{\rm tr}\{q_{2\mu}[G_0(\varepsilon)V\!G_0(\varepsilon)V\!G_0(\varepsilon)\!+\!
  G_0(\varepsilon){\cal V}^{(2)}G_0(\varepsilon)]\!\}.
\label{mpot}
\end{equation}
The solution of this equation has the form
\begin{equation}
{\cal V}^{(2)}({\bf r})=-\omega^2\sum_\mu\frac{\chi}{1+\chi\sigma_\mu}
     (-1)^\mu Q^{(2)}_{2\mu}q_{2-\mu}({\bf r}),
\label{mpot1}
\end{equation}
where
\begin{equation}
\sigma_\mu=\sum_{1,2}|(q_{2\mu})_{12}|^2\frac{\rho_1-\rho_2}
   {\varepsilon_1-\varepsilon_2}, \hspace{5mm}
Q^{(2)}_{2\mu}=\oint\limits_C\frac{d\varepsilon}{2\pi i}
     {\rm tr}\{q_{2\mu}G_0(\varepsilon)\ell_xG_0(\varepsilon)
               \ell_xG_0(\varepsilon)\}.
\end{equation}
The last quantity is the second correction to the nuclear quadrupole
moment due to rotation. Its explicit form is
\begin{equation}
Q^{(2)}_{2\mu}=\sum (q_{2\mu})_{12}\ell^x_{23}\ell^x_{31}
     \sum^2_{i=0}\!\hat P_i\left\{\frac{\rho_1}
{(\varepsilon_1-\varepsilon_2)(\varepsilon_1-\varepsilon_3)}\right\},
\label{qmom}
\end{equation}
where the permutation of indices $\nu\!=\!1,2,3$ by the operator 
$\hat P_i$ is subject to the rule $\nu\ \!{\rm mod}\ \!3\!=\nu$. It is 
obvious that the non-zero corrections have the components with 
$\mu=0,\pm 2$. The denominator in the sum (\ref{mpot1})
renormalizes the interaction strength. The straightforward
calculation of $\sigma_\mu$ and the use of Eq. (\ref{str}) with the
zero-order density matrix $\rho$ gives
$\chi/(1+\chi\sigma_\mu)=2\chi$. Combining (\ref{ctf}) with
(\ref{mpot1}), we have
\begin{equation}
\beta^{({\rm str})}_n=\frac{16\pi M^2\omega^2_0}
  {15\hbar\omega_z\Sigma_z}
   \sum_{\mu=0,\pm 2}Q^{(2)}_{2\mu}Q^{(2)}_{2-\mu}.
\label{stret}
\end{equation}

We can now calculate the two contributions to the parameter $\beta_n$
by summing over the quantum numbers $n_x$, $n_y$, and $n_z$.
The anisotropic oscillator potential allows to find an exact solution
of the problem. At first we find the corrections to the quadrupole moments
$$
Q^{(2)}_{20}=\sqrt{\frac{5}{64\pi}}
  \frac{\hbar\Sigma_z}{M\omega_0^2\omega_z(k^2-1)}
       (2k^4-15k^2+1),
$$
\begin{equation}
Q^{(2)}_{2\pm 2}=\sqrt{\frac{5}{128\pi}}
  \frac{\hbar\Sigma_z(1-5k^2)}{M\omega_0^2\omega_zk^2(k^2-1)}.
\end{equation}
Then by using (\ref{stret}) we obtain the contribution of the centrifugal
stretching effect
\begin{equation}
\beta^{({\rm str})}_n=\frac{\hbar\Sigma_z}{3\omega^3_zk^4(k^2-1)^2}
  (k^8-15k^6+76k^4-15k^2+1).
\end{equation}
Finally, after some fairly tedious calculations of the sum
(\ref{spur2}) we get
\begin{equation}
\beta^{({\rm sp})}_n=\frac{\hbar\Sigma_z}{2\omega^3_zk^4(k^2-1)^2}
   (k^8-10k^6-14k^4-10k^2+1).
\end{equation}
Adding the last two quantities gives us the parameter $\beta$ in the
normal phase:
\begin{equation}
\beta_n=\frac{5\Im_{\rm rig}}{6\omega^2_0}
         \frac{k^4-10k^2+1}{k^{2/3}(k^2+1)},
\label{spur3}
\end{equation}
if we use the following formula for the rigid body moment of inertia:
\begin{equation}
\Im_{\rm rig}=\frac{\hbar\Sigma_z}{\omega_zk^2}(k^2+1).
\label{moi1}
\end{equation}
Parameter $\beta_n$ is substantially reduced compared to $\beta_s$, 
$\beta_n\!\sim\!\hbar^4A^{7/3}/\varepsilon^3_F\!\sim
\beta_s({\rm SD})A^{-2/3}$. This can be explained by canceling main 
terms in the sums (\ref{spur2}) and (\ref{qmom}). That is exactly why
the corresponding values $\beta^{({\rm sp})}_n$ and $Q^{(2)}_{2\mu}$
are proportional to $\Sigma_z$. Such result is predictable because the
Hamiltonian $h_\omega$ for the anisotropic harmonic oscillator can be
diagonalized exactly \cite{R/S}. Its eigenstates are characterized by
the number of rotating bosons. To find $\beta^{({\rm sp})}_n$ and
$Q^{(2)}_{2\mu}$ we have to calculate first
the expectation values of the operators $\ell_x$ and $Q_{2\mu}$ in this
rotating basis. Then these quantities must be expanded in powers of
$\omega$. Because these operators are represented by
quadratic forms in the rotating bosons, their mean values and therefore
all the terms of the series are proportional to the linear combination
of $\Sigma_x$, $\Sigma_y$, and $\Sigma_z$.

Another peculiarity of the solution (\ref{spur3}) is that $\beta_n<0$
for the prolate nuclei with $1<b/a\!<3.15$, whereas $\beta_s$ is
always positive. The formal cause of this effect is a negative value
of $\beta^{({\rm sp})}_n$ and the inequality
$|\beta^{({\rm sp})}_n|\!>\!\beta^{({\rm str})}_n\!>\!0$ which is
fulfilled for the above indicated deformations. In the superfluid
phase, the term responsible for the rotation-quasiparticle
interaction may be also negative, but it never exceeds the
contribution of the pair modification effect (see Fig. 2a).

\setcounter{equation}{0}
\section{Analysis of experimental data}
\label{sec5}

We have shown in the preceding section that the second inertial
parameter ${\cal B}$ is negative in the superfluid phase and positive
in the normal one. The two limiting cases allow us to reconstruct the
${\cal B}(I)$ dependence for the parametrization (\ref{exp}) of a
rotational sequence with $(\pi\alpha)=(+0)$. Comparing the formulas
(\ref{ndbs}) or (\ref{spur1}) with (\ref{spur3}) we conclude that the
ratio ${\cal B}/{\cal A}$ has to change sign with increasing the spin
$I$ in a band and approach its limiting value
${\cal B}_n/{\cal A}_n\sim A^{-8/3}$ for high $I$.

The limiting ratio for a real nucleus can be obtained from
Eqs. (\ref{spur3}) and (\ref{moi1}) if we suppose that the r.m.s radius
and the deformation are exactly the same for neutron ($\nu$) and
proton ($\pi$) systems. The first condition implies that oscillator
frequencies of neutron and proton potentials satisfy the relation
$\omega_{0\tau}=\omega_0(2A_\tau/A)$ ($\tau=\pi,\nu$ and
$A_\tau$ is the number of nucleons of a given type). The second one
results in the identical ratio of the frequencies along the principal axes
for the both potentials:
\begin{equation}
\omega_{x\tau}:\omega_{y\tau}:\omega_{z\tau}=m:m:l.
\end{equation}
For integers $m$ and $l$, the states with the same number of quanta
${\cal N}_{ml}=m(n_x+n_y)+ln_z$ form a deformed shell. Assuming that,
for a given number of nucleons $A_\tau$, all the ${\cal N}_{ml}$-shells
are filled, one can express the sum $\Sigma_{z\tau}$ in the form
\begin{equation}
\omega_{z\tau}\Sigma_{z\tau}=
\omega_{0\tau}(\Sigma_{x\tau}\Sigma_{y\tau}\Sigma_{z\tau})^{1/3}=
\omega_{0\tau}\left(\frac{A_\tau^4}{32}\right)^{1/3}.
\end{equation}
The above formulas allow us to derive the ratio
${\cal B}/{\cal A}$ for a nucleus consisting of $Z$ protons and $N$
neutrons in the normal state:
\begin{equation}
\frac{{\cal B}_n}{{\cal A}_n}\!=\!-3.205\frac{(k^4\!-\!10k^2\!+\!1)k^{2/3}}
   {(k^2+1)^3A^{8/3}}\!\left[\left(\frac{Z}{A}\right)^{\!1/3}\!+
  \left(\frac{N}{A}\right)^{\!1/3}\right].
\label{ratio}
\end{equation}
This result holds for a nucleus with an arbitrary deformation
$k=b/a=m/l$.

We concentrate first on SD bands. Most of them are not connected
to lower-lying states of known excitation energy, spin, and parity.
Thus, their exit spins $I_0$ are unknown.
Tentative spin assignment is used to take advantage
of the formulas (\ref{para}) to find the experimental ratio
${\cal B}/{\cal A}$. To analyze this quantity, we will take into
account two basic ingredients: shell gaps, which stabilize the shape,
and intruder orbitals involved in the alignment. The
nucleon-configuration assignment of a band is generally based only
on the behavior of the dynamic moment of inertia and the quadrupole
moment in a given band. The last quantity,
\begin{equation}
 Q_0=6.05\cdot 10^{-3}A^{2/3}\ \frac{k^2-1}{k^{2/3}}eb,
\label{qud}
\end{equation}
remains remarkably constant as a function of spin within a band. This
proves that the deformation $k$ remains practically unchanged as $I$
increases. We use the experimentally observed value of $Q_0$ to
find the axis ratio $b/a$, which is required for calculation of the quantity
(\ref{ratio}).

The ratio ${\cal B}/{\cal A}$ extracted from the measured energy of
$\gamma$-transitions in the four SD bands of the $A=150$ mass region
are shown in Fig. 4. The parity
and the signature of these bands is assumed to be $(+,0)$. We also
use the adopted spins for their lowest levels. The band $^{152}$Dy(1)
belongs to the doubly-magic nucleus with the proton $Z=66$ and the
neutron $N=86$ gaps in single-particle spectra at the same deformation
\cite{BeN}. The gaps decrease a level density and considerably reduce
the neutron and proton pairing correlations. There is no direct
experimental indication of pairing correlations in this band. The
theoretical calculations \cite{Aou} show that their inclusion leads to
a better description of the kinematic and dynamic moments of inertia,
pairing correlations being more important at the low spin range. The
plot shows that there are two distinct regions in the variation of
the ratio ${\cal B}/{\cal A}$ versus $I$. The lower part of the band
exhibits a sharp increase of this quantity. It then changes sign at
the spin $I_c=36$ and approaches the plateau value of (\ref{ratio})
at the top of the band. Such behavior of the ratio apparently shows that
the static pairing correlations of neutrons and protons are
quenched simultaneously. This fact also proves that most part of the band
belongs to the normal phase.

The band $^{144}$Gd(1) is one of the few examples of SD bands which
exhibits backbending. The $\pi6^2$ pair alignment opens up the
proton shell gap  $Z=64$ at the same deformation as the neutron shell
gap $N=80$. Thus, above the backbending this band becomes similar
to the doubly-magic $^{152}$Dy(1) except the gap $Z=64$ is
less pronounced than the one at $Z=66$. Besides, the neutron gap
$N=80$ is somewhat smaller than $N=86$. These factors
enhance a level density and favor pairing correlations. As seen in
Fig. 4 the behavior of ${\cal B}/{\cal A}$ for this band in the
low-$I$ region is the same as that for $^{152}$Dy(1) if we scale the
axis of ordinates by the factor two. Accordingly, the critical
value is somewhat larger, $I_c=38$.

The features observed at low spins in the dynamic moment of inertia
of the band $^{150}$Gd(1) have been explained in terms of consecutive
alignments of the $\nu7^2$ and $\pi6^2$ pairs \cite{BeN}. For the
configuration $\pi6^2\nu7^2$ all levels below the $Z=64$ and $N=86$
shell gaps are occupied. The former is found at slightly smaller
deformation than the latter. This factor diminishes the neutron gap and
enhances neutron pairing. The Woods-Saxon \cite{BeN} and the
relativistic mean field \cite{Afan} calculations make evident that
static proton and neutron pairing exist at low spins, $I<48$
($\hbar\omega<0.55$ MeV). In addition, the full self-consistent
HFB calculations with the particle number
projection \cite{Aou} show that the effect of pairing on the
moments of inertia in $^{150}$Gd(1) is about twice as important as
in $^{152}$Dy(1). It is apparent from Fig. 4 that the static pairing
correlations  in the band $^{150}$Gd(1) are stronger than in
$^{144}$Gd(1). One should expect even more stronger pairing
correlations in the newly discovered prolate deformed band
$^{154}$Er(2) \cite{Lag}, since the proton Fermi level at $Z=68$ lies
in the region of a high level density above the $Z=66$ gap.
The experimental ${\cal B}/{\cal A}$ dependence of Fig. 4 is
consistent with this prediction. It is seen that this ratio does not
exceed the value $-5\times10^{-5}$ and does not show the
plateau.\footnote{It is worth mentioning that the nonaxial band 
$^{154}$Er(1) demonstrates the ${\cal B}/{\cal A}$ dependence with 
the critical spin $I_c=31$ and a long plateau.} The bump
seen at $I=44$ ($\omega=0.57$ MeV) can be attributed to the
alignment of a pair of $i_{13/2}$ protons in agreement with the
calculations of Ref. \cite{BeN}. Thus, the plots of Fig. 4 show the
correlation of the spin dependence of the ratio ${\cal B}/{\cal A}$
with the level density near the Fermi surface: the higher the level
density, the stronger pairing correlations, and the less marked the
plateau.

The high deformed (HD) bands in the $A=190$ mass region are related
to the $Z=80$ and $N=112$ shell gaps. Most of these bands have
similar values of $\Im^{(2)}$ that exhibits a smooth rise as a function of
rotational frequency. This rise is attributed to the gradual alignment,
in the presence of static pairing correlations, of $i_{13/2}$ protons
and $j_{15/2}$ neutrons. The calculations with pairing are able to
reproduce the general trend seen in experiment. The bands
$^{194}$Hg(1) and $^{194}$Pb(1) are of central importance because
their spins, parities, and excitation energies are known \cite{Hac,Hau}.
The plots of Fig. 5 for these bands demonstrate the gradual rise of
the ${\cal B}/{\cal A}$ ratio that confirms the
presence of static pairing correlations.

Now we turn our attention to ND bands. There are several bands of
isotopes Er, Yb, and Hf in which the static neutron pairing gaps are
predicted to collapse. However, the proton system still has strong
pairing correlations. Accordingly, the
${\cal B}/{\cal A}$ vs. $I$ plot for these bands exhibits a
sharp rise, but does not approach the plateau.
The yrast band of $^{84}$Zr is an exception. Because
protons and neutrons occupy in this nucleus similar orbitals near the
Fermi surface, quasiparticle alignments and the elimination of
pairing gaps occur at similar spin. Besides, the deformed shell gaps
at $Z=N=38$ and the low moment of inertia favor the transition to the
normal phase. Combination of these factors makes the pairing
properties of the ND band of $^{84}$Zr similar to those of SD bands
in the $A=150$ mass region. Fig. 6 plots, as a function of
spin $I$, the ${\cal B}/{\cal A}$ ratios for the ND and SD bands of
this nucleus determined from the data of Refs. \cite{Pric} and
\cite{Jin}, respectively. In the SD band $^{84}$Zr(1), static pairing
correlations are quenched completely due to the high rotational
frequency. This inference is supported by the coincidence of
experimental points with the plateau ${\cal B}_n/{\cal A}_n$. It also
becomes apparent from this figure that at high spins the experimental
ratio for the ND band reaches the same plateau.
The difference in the limiting value of ${\cal B}_n/{\cal A}_n$ due to
the difference in deformations ($\beta=0.43$ and 0.55 for the ND and
SD bands, respectively) is insignificant. The low spin part ($I<18$)
of the ND band is compatible with the transitional nature of the
$\gamma$-soft nucleus: small $\beta$ and a noticeable triaxiality.
The alignment of the two $g_{9/2}$ quasiprotons and the subsequent
alignment of the two $g_{9/2}$ quasineutrons are clearly seen in
Fig. 6 as the humps A and B. Beyond the second alignment, a
striking change in deformation occurs in the interval of spins
$I=18-22$. After the spin $I=24$ the rotational behavior is
compatible with the rigid rotation of a high deformed axially
symmetric nucleus \cite{Pric}.

The characteristic behavior of the ratio ${\cal B}/{\cal A}$ with the
critical spin $I_c$ and the pronounced plateau have also been found
in the SD bands of the $A=150$ nuclei having configurations different
from $(+,0)$. The bands $^{152}$Tb(2) [the $\pi[301]1/2$ hole in the
$^{152}$Dy SD core] and $^{153}$Ho(3) [the $^{152}$Dy SD core coupled
to the $\pi[523]7/2$ orbital] show the ${\cal B}/{\cal A}$ dependence
similar to that of the $^{152}$Dy(1) band. The pair of identical bands
$^{150}$Gd(2) and $^{151}$Tb(1) have the dependence similar to
one of $^{150}$Gd(1). All these bands have somewhat higher values of
${\cal B}/{\cal A}$ than do their $(+,0)$ counterparts. This proves that an
odd nucleon or a particle-hole excitation reduces pairing correlations
due to the blocking effect. Such phenomenon is characteristic of the
static pairing regime \cite{Shi}. The SD bands from different mass
regions, $^{132}$Ce(1), $^{133}$Ce(1), and $^{60}$Zn(1), exhibit the
same behavior of ${\cal B}/{\cal A}$. A strong configuration
dependent effect is observed in the bands where the odd neutron is
placed in the $j_{15/2}$ intruder orbitals. For such bands, the
${\cal B}/{\cal A}$ ratio is positive for all spins. The examples include
the bands $^{149}$Gd(1) (configuration $\pi6^2\nu7^1$), $^{151}$Dy(1)
($\pi6^4\nu7^1$), and $^{153}$Dy(1) ($\pi6^4\nu7^3$). A
single-particle degree of freedom seems to destroy the typical
behavior of the ratio ${\cal B}/{\cal A}$. More efforts are needed
to explain this interesting feature.

These numerous examples prove the universality of the transition from
the superfluid to the normal phase for SD and ND bands. This
universality can be represented, according to Ref.\ \cite{Pav}, by the
effective rotational Hamiltonian,
\begin{equation}
H_{\rm eff}={\rm a}{\bf I}^2 + (I/I_c-1){\rm b}{\bf I}^4+
        {\rm c}{\bf I}^6,
\label{eff}
\end{equation}
which describes the states of the $(+,0)$ band in the transition region.
The parameters {\rm a}, {\rm b}, {\rm c}, and the critical spin $I_c$
are the subjects of the microscopic theory, which has to take
into account static, dynamic, and uniform pairing. Incorporating the
critical spin, which can be found from the experimental plot
${\cal B}/{\cal A}$ vs. $I$, this concept of the superfluid-to-normal
transition is free from ambiguities characteristic of the approach
based on a change of the single-particle spectra \cite{Gar,Oliv}.

Using the results of our analysis we can now explain why some SD bands
have extremely regular rotational spectra. Figure 4
shows that the most part of the bands $^{152}$Dy(1) and $^{144}$Gd(1)
belongs to the plateau with the ratio of the inertial parameters
${\cal B}/{\cal A}\sim 10^{-6}$. So does the whole of the SD band
$^{84}$Zr(1), for which this ratio is $10^{-5}$. The plateau is the
manifestation of the normal phase with the anomalous small ratio
(\ref{ratio}), ${\cal B}_n/{\cal A}_n\sim A^{-8/3}$. The above values
for the SD bands agree with this estimation. Therefore the extreme
regularity is explained by quenching the static pairing correlations
in the lower parts of these bands. On the contrary, the
transition in the yrast band of $^{84}$Zr occurs in its upper part.
Accordingly, the top of the band has the same properties. It is
important to note that the bands in which proton and neutron pairing
gaps are present [$^{154}$Er(2) and all the SD bands in the $A=190$
mass region] and the bands with proton pairing alone [$^{168}$Yb(yr)
and $^{186}$Hf(yr)] are regular to a lesser extent.

We should mention one more feature which requires further
investigation. The down-sloping of the ${\cal B}/{\cal A}$ dependence
is observed  at the top of $^{152}$Dy(1), $^{84}$Zr(1), and other SD
bands with extremely high spins. Because the quantity
${\cal B}_n/{\cal A}_n$ is a decreasing function of the deformation
$k$, it would be natural to explain this feature by the increase of the
nuclear elongation due to the enormous centrifugal stretching at the
end of these bands.

\setcounter{equation}{0}
\section{Conclusion}
\label{sec6}

Despite the vast amount of data collected and various theoretical
interpretations suggested, a detailed understanding of many properties
of SD bands has yet to be achieved. Pairing correlations are just one
example of such properties.
The presence of static pairing in SD bands is usually established by
studying the behavior of the dynamic moment of inertia $\Im^{(2)}$ as
a function of the rotational frequency $\omega$. A band crossing
associated with a quasiparticle alignment leads to a impressive
decrease in $\Im^{(2)}$ with $\omega$ or a hump in this dependence.
This gives an indication that static pairing correlations are present in
that part of a band where such irregularities occur.

In this paper, the investigation of pairing correlations is based on
the spin dependence of the second inertial parameter ${\cal B}$.
This quantity, which is proportional to the difference
$\Im^{(1)}-\Im^{(2)}$ in the high-$I$ limit, turned out to be a more
sensitive measure of the change in pairing correlations
than $\Im^{(2)}$. The new method requires
spin-signature assignments of the band states. However, it
gives more definite information about the superfluid-to-normal
transition in a band. The most important results obtained in this
paper can be summarized as follows:

(i) The exact semiclassical expression for the second inertial
parameter in the superfluid phase has been found by taking into
account the effect of rotation on the Cooper pairs in the
gauge-invariant form. The presence of nonuniform pairing reduces the
nonadiabatic effect of rotation. Its influence increases strongly at
superdeformation. The nonuniform pairing allows one to find correctly
the interesting physical limits for the second inertial parameter.

(ii) The limit of zero static pairing is of special interest. It
permits the function ${\cal B}(I)$ to be reconstructed by
interpolating between the values of ${\cal B}$ in the superfluid and
normal phases. The anisotropic oscillator model calculations show
that there are two distinct regions in the variation of the ratio
${\cal B}/{\cal A}$ with $I$. The lower part of a band is
characterized by a gradual decrease of pairing. Accordingly, being
negative the ratio ${\cal B}/{\cal A}$ exhibits a sharp increase. It
then changes the sign at the spin $I_c$ and approaches the positive
value characteristic of the normal phase. The critical point
$I_c$, ${\cal B}(I_c)=0$, is a signature of the superfluid-to-normal
transition. The transition manifests itself in the modification of
the rotational spectrum of a band.

(iii) The experimental spin dependence of ${\cal B}/{\cal A}$ indicates
an agreement with the theoretical prediction and demonstrates the
universality of the transition to the normal state. This agreement is
not a trivial fact because our calculations are based on the simplest
model of a nuclear potential and do not take into account pairing
fluctuations in the normal phase. Nevertheless, the agreement is not
accidental because the universal dependence of ${\cal B}/{\cal A}$ on
$I$ has been observed  for a large number of SD bands and some ND
ones.

(iv) The universal spin dependence of ${\cal B}/{\cal A}$ explains the
extreme regularity of some SD bands. The characteristic feature of
this dependence is the pronounced plateau in the upper part of a SD
band ($I>I_c$) corresponding to the normal phase. The calculated
ratio in this part of a band is extremely small,
${\cal B}/{\cal A}\sim A^{-8/3}$. Thus, the closer the critical point
$I_c$ to the exit spin $I_0$, the more regular its rotational
spectrum. The spectacular examples are the bands $^{144}$Gd(1)
and $^{152}$Dy(1) having ${\cal B}/{\cal A}\sim 10^{-6}$.

(v) Some new features have been observed in the upper parts of
SD bands. The investigation of this region, which is free from pairing
correlations, is extremely important for our understanding of the
microscopic structure at the superdeformed minimum.

\section*{Acknowledgments}

This work is supported in part by the Russian Foundation for Basic
Research, project no. 00-15-96590.

\renewcommand{\theequation}{\Alph{section}.\arabic{equation}}
\renewcommand{\thesection}{Appendix \Alph{section}.}
\setcounter{section}{0}
\setcounter{equation}{0}
\section{Solution of the integral equations for the \\nonuniform
     pairing field}

The effect of rotation on pairing correlations is described by the first,
$\Delta^{(1)}({\bf r})$, and the second, $\Delta^{(2)}({\bf r})$,
corrections to the pairing field, which enter into Eq. (\ref{fin}). We
have seen in Section \ref{sec2} that the integral equations that have
to be solved are of the general form (\ref{even}) and (\ref{odd}) for
even and odd corrections, respectively. It is convenient to introduce
into these equations the operator $\dot\ell_x$ that is a function of
the space coordinates only.

We start by considering the equation for $\Delta^{(1)}$. Using the
relation $\hbar\dot\ell^x_{12}=i(p_1-p_2)\ell^x_{12}$ we get from
(\ref{odd}) for $i=0$
\begin{equation}
\sum_{1,2}\int\limits_{C'}\frac{dp}{2\pi}\frac{1}
{({\bf p}_1^2+\Delta^2)({\bf p}_2^2+\Delta^2)}
[2i\Delta\hbar\omega\dot\ell^x_{12}+(p_1-p_2)^2\Delta^{(1)}_{12}]
\varphi_1({\bf r})\varphi^*_2({\bf r'})=0.
\label{odd1}
\end{equation}
The equation is satisfied if we assume that
\begin{equation}
\Delta^{(1)}({\bf r})=-i\frac{\hbar\omega}{2\Delta}D_1\dot\ell_x,
\label{odds}
\end{equation}
where the amplitude $D_1$ is determined after substituting (\ref{odds})
into Eq. (\ref{odd1}) and integrating over $\bf r$:
\begin{equation}
D_1\!=\!
4\Delta^2\sum_{1,2}\!|\dot\ell^x_{12}|^2\!\int\limits_{C'}\!\frac{dp}{2\pi}
\frac{1}{({\bf p}_1^2+\Delta^2)({\bf p}_2^2+\Delta^2)}\Bigg/\!
\sum_{1,2}\!|\dot\ell^x_{12}|^2\!\int\limits_{C'}\!\frac{dp}{2\pi}
\frac{(p_1-p_2)^2}{({\bf p}_1^2+\Delta^2)({\bf p}_2^2+\Delta^2)}.
\end{equation}
The solution transforms to
\begin{equation}
D_1=\frac{\sum_{1,2}|\dot\ell^x_{12}|^2g(x_{12})
\delta(\varepsilon_1\!\!-\!\varepsilon_F\!)}
{\sum_{1,2}|\dot\ell^x_{12}|^2x^2_{12}g(x_{12})
\delta(\varepsilon_1\!\!-\!\varepsilon_F\!)}
\label{amp1}
\end{equation}
in the semiclassical approximation. The summation over the state 1
is to be understood as the integration over its quantum numbers. For
the anisotropic oscillator potential, the amplitude can be expressed
in a simple analytical form
\begin{equation}
D_1=\frac{g(\nu_1)+g(\nu_2)}{\nu^2_1g(\nu_1)+\nu^2_2g(\nu_2)}.
\label{amh}
\end{equation}
The function $g(x)$ and the parameters $\nu_1$ and $\nu_2$ are
determined by Eqs. (\ref{fuhg}) and (\ref{osc}), respectively.

The equation for $\Delta^{(2)}({\bf r})$ after introducing $\dot\ell_x$
becomes
$$
\sum_{1,2,3}\left[2\hbar^2\omega^2\dot\ell^x_{12}\dot\ell^x_{23}
\int\limits_{C'}\frac{dp}{2\pi}\frac{{\cal Q}_3(p,p_1,p_2,p_3)}
{({\bf p}_1^2+\Delta^2)({\bf p}_2^2+\Delta^2)({\bf p}_3^2+\Delta^2)}\right.
$$
\begin{equation}
\left.-\Delta^{(2)}_{12}\delta_{23}\int\limits_{C'}\frac{dp}{2\pi}
\frac{(p_1-p_2)^2+4\Delta^2}
{({\bf p}_1^2+\Delta^2)({\bf p}_2^2+\Delta^2)}\right]
\varphi_1({\bf r})\varphi^*_2({\bf r'})=0,
\label{even1}
\end{equation}
where the polynomial function ${\cal Q}_3$ of the third order in 
$(p,p_\nu)$ depends also on the amplitude $D_1$. We try to solve this
equation by making substitution
\begin{equation}
\Delta^{(2)}({\bf r})=\frac{\hbar^2\omega^2}{4\Delta^3}D_2\dot\ell^2_x.
\label{evns}
\end{equation}
Applying the same procedure as before, one can find the amplitude
$D_2$ in the semiclassical approximation
\begin{equation}
D_2=\frac{\sum_{1,2,3}\dot\ell^x_{12}\dot\ell^x_{23}(\dot\ell^2_x)_{31}
\phi (x_{12},x_{13},x_{23})\delta(\varepsilon_1\!\!-\!\varepsilon_F\!)}
{\sum_{1,2,3}|(\dot\ell^2_x)_{12}|^2h(x_{12})
\delta(\varepsilon_1\!\!-\!\varepsilon_F\!)},
\label{amp2}
\end{equation}
where the function $h(x)$ is determined by (\ref{fuhg}) and that of
$\phi$ has the form
$$
\phi (x,y,z)=\frac{1}{2x^2y^2z^2}[-xy(1-D_1x^2)(1+xy-D_1z^2)g(x)
$$
\begin{equation}
 +y^2(1-D_1x^2-D_1z^2)h(y)
    -yz(1-D_1z^2)(1+yz-D_1x^2)g(z)].
\end{equation}
Their symmetry properties are
\begin{equation}
\phi (z,y,x)=\phi (x,y,z), \hspace{7mm} \phi (-x,-y,-z)=\phi (x,y,z).
\end{equation}
In the oscillator potential, the sum over the states 2 and 3 in the
numerator of (\ref{amp2}) comprises 16 terms including four with two
close transitions, four with two distant transitions, and eight terms with
one close and one distant transitions. Performing summation in the
semiclassical approximation in the numerator and in the denominator
of (\ref{amp2}), we find
$$
D_2=[4\phi(\nu_1,\nu_1-\nu_2,-\nu_2)+4\phi(\nu_1,\nu_1+\nu_2,\nu_2)
$$
\begin{equation}
+\phi(\nu_1,2\nu_1,\nu_1) +\phi(\nu_2,2\nu_2,\nu_2)
+4\phi(\nu_1,0,\!-\nu_1)
   +4\phi(\nu_2,0,\!-\nu_2)]{\cal H}^{-1}(\nu_1,\nu_2),
\label{amh1}
\end{equation}
where
\begin{equation}
{\cal H}(\nu_1,\nu_2)=8+4h(\nu_1-\nu_2)+4h(\nu_1+\nu_2)
     +h(2\nu_1)+h(2\nu_2).
\label{amh2}
\end{equation}

For the monopole pairing interaction, the pairing field is uniform and
the first correction $\Delta^{(1)}$ vanishes. The coordinate
independent solution for the correction $\Delta^{(2)}$  can be found
from Eq. (\ref{even1}) after averaging over $\bf r$. The resulting
expression may be written in terms of the kinematic moment of
inertia,
\begin{equation}
\Delta^{(2)}=\frac{\omega^2}{2\rho_F}
        \frac{\partial\Im^{(1)}}{\partial\Delta}, \hspace{7mm}
\Im^{(1)}=\sum_{1,2}|\ell^x_{12}|^2[1-g(x_{12})]
\delta(\varepsilon_1\!\!-\!\varepsilon_F\!),
\end{equation}
in agreement with the result obtained by Marshalek \cite{Mar}. From
the theoretical viewpoint, this solution is not correct because it
violates the current conservation.

\renewcommand{\theequation}{\Alph{section}.\arabic{equation}}
\renewcommand{\thesection}{Appendix \Alph{section}.}
\setcounter{section}{1}
\setcounter{equation}{0}
\section{Calculation of integrals}

In this Appendix we give a brief outline of the technique used in
calculation of the integrals which are necessary for obtaining the
function $F$ (\ref{fun}) and for the solution of the integral equations
(\ref{odd1}) and (\ref{even1}). All the relevant integrals can be done
exactly with the method proposed by Feynman in the quantum
electrodynamics \cite{Fyn}. The method is based on the identity
\begin{equation}
\frac{1}{a_1a_2...a_n}\!=\!(n-1)!\!\int\limits_0^1dt_1
  \int\limits_0^{t_1}dt_2...\!\!\!\int\limits_0^{t_{n-2}}\!\frac{dt_{n-1}}
   {[a_1t_{n-1}\!+\!a_2(t_{n-2}\!-\!t_{n-1})\!+...+\!a_n(1\!-\!t_1)]^n},
\label{feyn}
\end{equation}
which is proved by a direct calculation.

The simplest integral is the one involved in the sum (\ref{sum1}). It is solved
by using (\ref{feyn}) as follows:
$$
J_1=\int\frac{d{\bf p}_1}{2\pi}\frac{1}
  {({\bf p}_1^2+\Delta^2)({\bf p}_2^2+\Delta^2)}
$$
\begin{equation}
=\int\limits_0^1dt\int\frac{d{\bf p}_1}{2\pi}\frac{1}
  {[{\bf p}_1^2+p_{12}^2Q(t)]^2}=
\frac{1}{2p^2_{12}}\int\limits_0^1\frac{dt}{Q(t)}=
\frac{1}{2\Delta^2}g\!\left(\frac{p_{12}}{2\Delta}\right),
\end{equation}
where
$$
Q(t)=-t^2+t+\delta^2,\hspace{4mm} \delta=\Delta/p_{12},
\hspace{4mm} p_{12}=p_1-p_2.
$$

Four integrals appear in the first sum of Eq. (\ref{even1}). All those
are of the same type. As an example, we consider the term
proportional to the square of the amplitude $D_1$. The relevant
integral is
$$
J_2=\Delta\int\frac{d{\bf p}_1}{2\pi}\frac{{\cal Q}_2(p;p_1,p_2,p_3)}
{({\bf p}_1^2+\Delta^2)({\bf p}_2^2+\Delta^2)({\bf p}_3^2+\Delta^2)}
$$
\begin{equation}
=\Delta\int\limits_0^1dt_1\int\limits_0^{t_1}dt_2
   \int\frac{d{\bf p}_1}{2\pi}
      \frac{{\cal Q}_2(p;p_1+(t_1-t_2)p_{12}+(1-t_1)p_{13})}
  {[{\bf p}_1^2+p_{13}^2Q(t_1,t_2)]^3},
\end{equation}
where ${\cal Q}_2=p^2+p_1p_2-p_1p_3+p_2p_3+\Delta^2$,
$$
Q(t_1,t_2)=-[(1-c^2)t_1+c^2t_2-1]^2-(1-c^2)t_1-c^2t_2+1+\delta^2,
$$
and $ c=p_{12}/p_{13}$. By making the substitution
$u_1=(1-c)t_1+ct_2$, $u_2=t_2$, we get after integration over
$d{\bf p}_1$
\begin{equation}
J_2=\frac{\Delta}{2p_{13}p_{23}}
\left\{\int\limits_0^{1-c}\!du_1\int\limits_0^{u_1}\!du_2+
 \int\limits_{1-c}^1\!du_1\!\!\!
    \int\limits_{(u_1-1+c)/c}^{u_1}\!\!\!\!\!du_2\right\}
\frac{(1\!-\!c)u_1\!-\!cu_2\!-\!1\!+\!c\!-\!2\delta^2}{[cu_2+Q_1(u_1)]^2},
\end{equation}
where $Q_1(u_1)=-u^2_1+(1-c)u_1+\delta^2$. Straightforward calculation
of these integrals gives
\begin{equation}
J_2=\frac{1}{2\Delta x_{13}}[x_{12}g(x_{12})+x_{23}g(x_{23})].
\end{equation}

Finally, let us consider the integral (\ref{subst}). It is solved in the
same manner as the preceding ones. After using the identity
(\ref{feyn}) and integration over $d{\bf p}_1$, the substitution
$t_1=[u_1-(d-c)u_2-cu_3]/(1-d)$, $t_2=u_2$, $t_3=u_3$, where
$c=p_{12}/p_{14}$, $d=p_{13}/p_{14}$, leads to four triple integrals
which may be solved without a problem.

\newpage
\begin{center}
FIGURE CAPTIONS
\end{center}

\vspace{8mm}
Fig. 1.
Relative deviation of energies $E$ in the superdeformed band
$^{194}$Pb(1) as compared with the ground state band of $^{238}$U
and the ground vibrational state band of the H$_2$ molecule. The
deviation is calculated from the formula $(E-E_{\rm{rig}})/E_{\rm{rig}}$,
where $E_{\rm{rig}}={\cal A}I(I+1)$, and the parameters ${\cal A}$ are
found from the energies $E_\gamma(4)$ of the $6\to 4$ transitions.
The experimental data are taken from Refs. \cite{Fir} and \cite{Dab}.

\vspace{5mm}
Fig.\ 2.
Comparison of the functions relevant to the second inertial parameter for
ND bands. Part (a) represents the function $\Phi_n$ (\ref{sdbs})
(solid line), $\Phi_{nc}$ (\ref{ndf}) (dashed line), and their parts,
the functions $\cal F$ (\ref{nsf}) (dotted line) and
${\cal F}_{\rm c}$ (\ref{fcl}) (dash-dotted line) which describe
the effect of rotation-quasiparticle interaction. Part (b) is for the
limit of close transitions. The function $\Phi_{nc}$ (solid line)
and its constituents ${\cal F}_c$ (dashed line) and
$2D^2_2{\cal H}_c$ (dotted line) are shown for $\Delta^{(1)}=0$.
The axis of abscissas shows the dimensionless quantity $\nu_1$
corresponding to the close transitions, while that for the distant
ones is fixed by the representative value $\nu_2=10$ for all the plots.

\vspace{5mm}
Fig.\ 3.
Plot of the function $\Phi$, to which the second inertial parameter
for SD bands (\ref{spur1}) is proportional, against the dimensionless
quantity $\xi$ for the axis ratio $b/a=2$. The solid and dashed lines
correspond to the exact value and the limit of uniform pairing,
respectively. The abscissa scale must be multiplied by a factor of
approximately 7.7 for nuclei in the $A\sim150$
mass region in order to obtain the gap energy in MeV.

\vspace{5mm}
Fig.\ 4.
Ratio ${\cal B}/{\cal A}$ versus spin for some SD bands of the
$A=150$ region. Expressions (\ref{para}) are used to extract this ratio
from experimental data taken from Refs. \cite{Fir} and \cite{Lag}. The
error bars (if they are greater than symbol sizes) include only the
uncertainties in $\gamma$-ray energies. The uncertainties in the spin
assignment are immaterial for all bands [with the exception of
$^{152}$Dy(1)], since the spin variation in 2$\hbar$ would merely
shift the curves along the abscissa. The experimental points for
the band $^{144}$Gd(1) are shown above the $\pi i_{13/2}$
backbending. The solid straight line is the ratio
${\cal B}_n/{\cal A}_n$ (\ref{ratio}) for the normal phase with the
deformation $b/a$ found from the quadrupole moment (\ref{qud}).

\vspace{5mm}
Fig.\ 5.
The same as in Fig. 4 for the two SD bands of the $A=190$ region.
Experimental data are taken from Refs.  \cite{Fir,Hac,Hau}.

\vspace{5mm}
Fig.\ 6.
Ratio ${\cal B}/{\cal A}$ versus spin for the yrast ND band (solid
circles) and the SD band (open circles) of $^{84}$Zr. Experimental
data are taken from Refs. \cite{Pric} and \cite{Jin}. The solid
straight line is the ratio ${\cal B}_n/{\cal A}_n$ (\ref{ratio})
relevant to the deformation of the SD band.

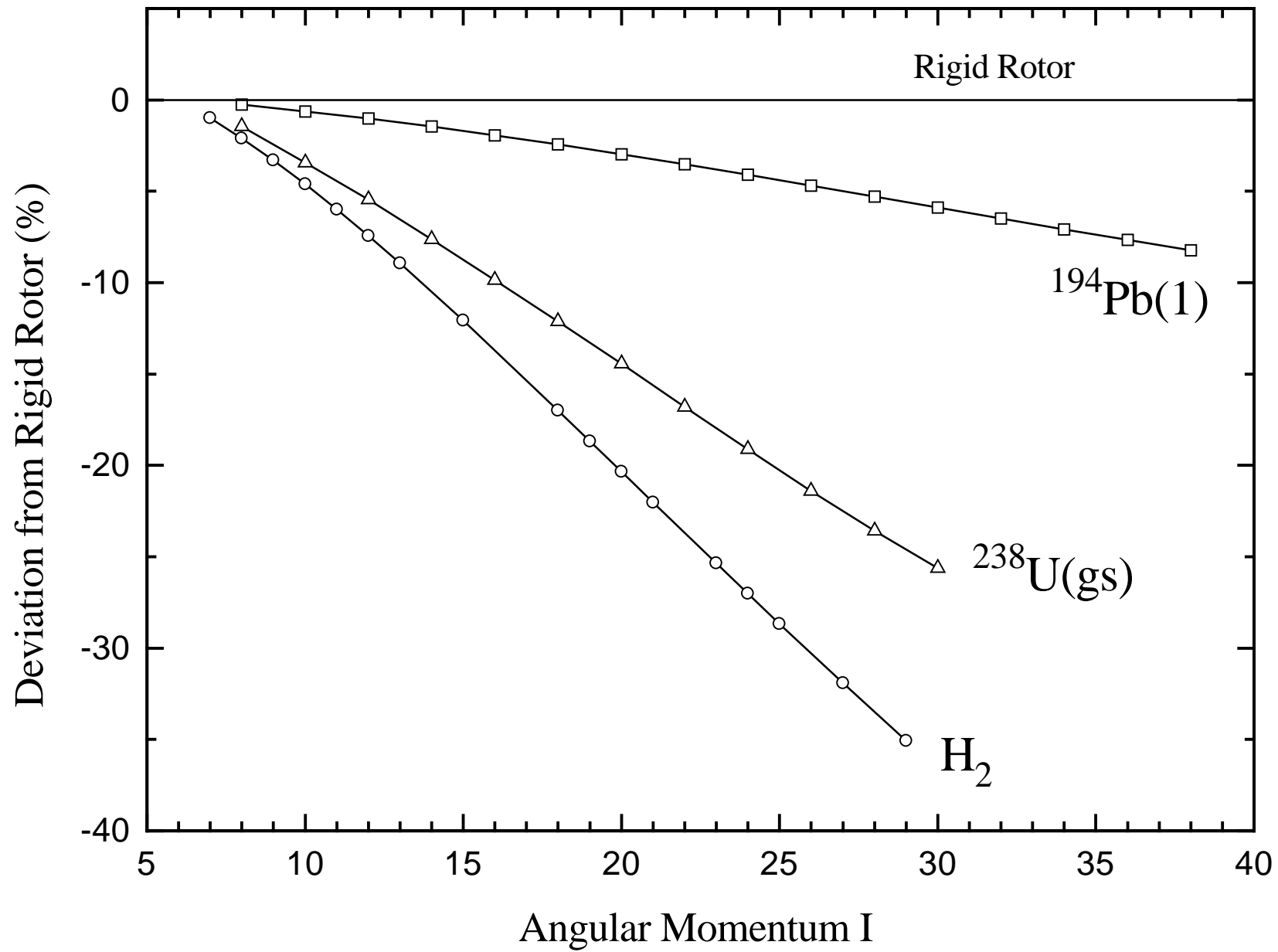
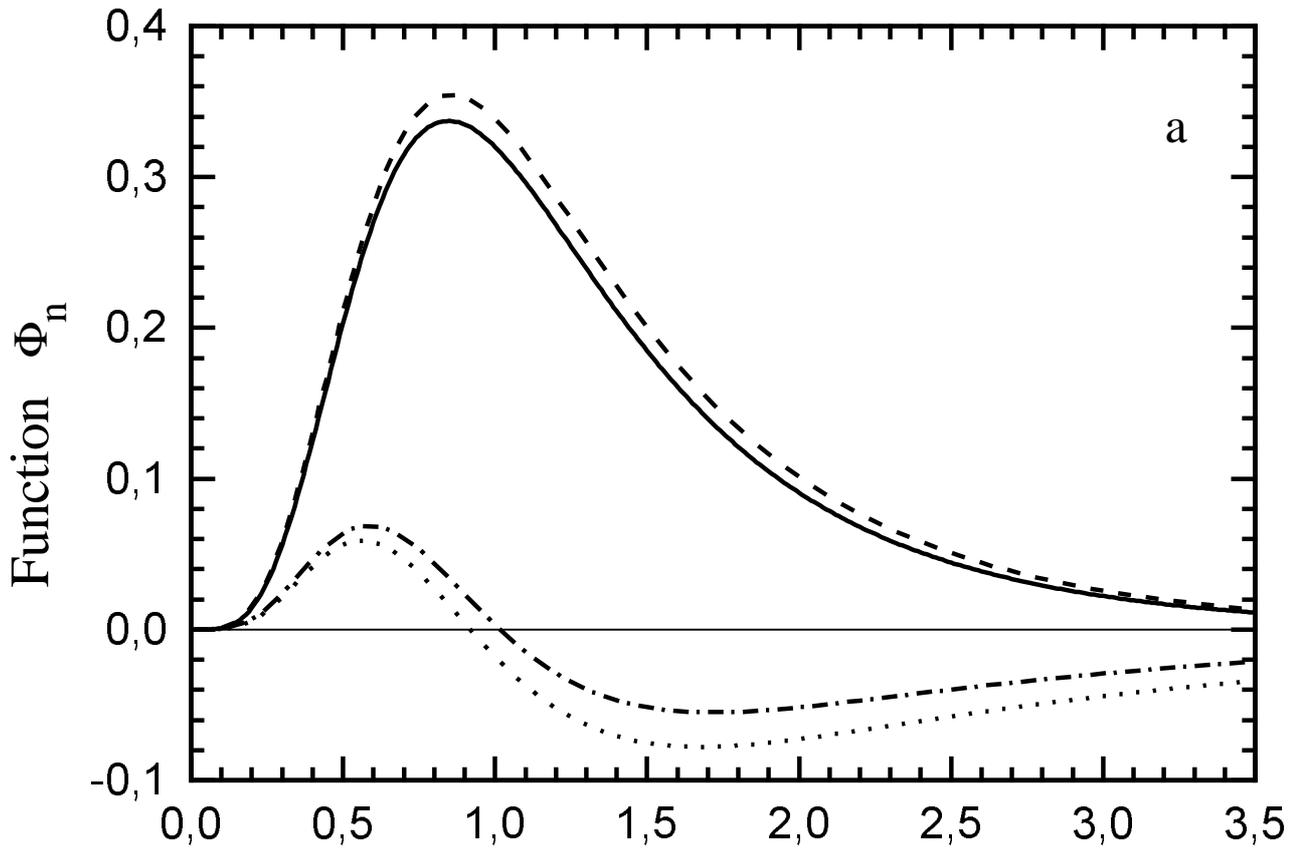
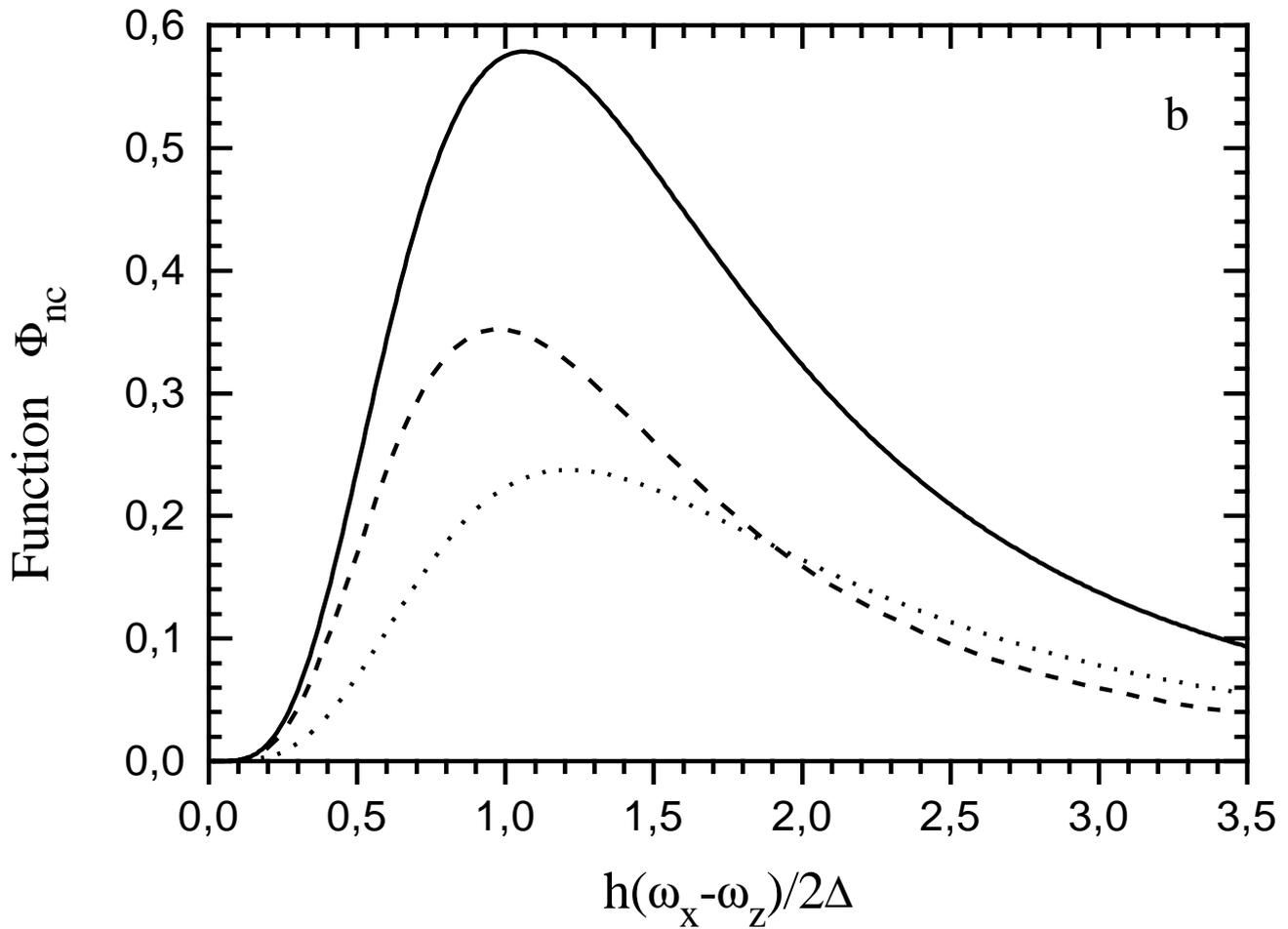
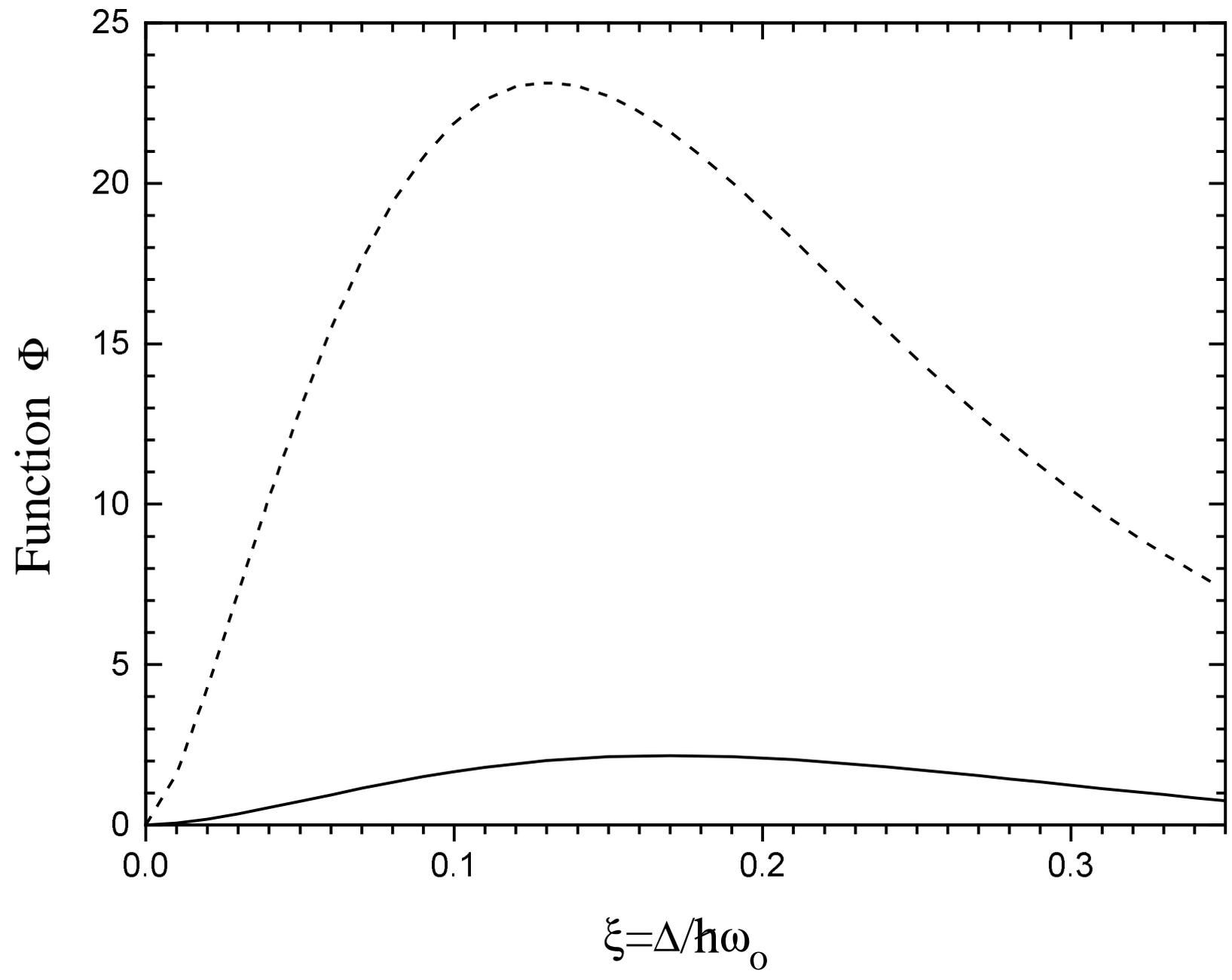
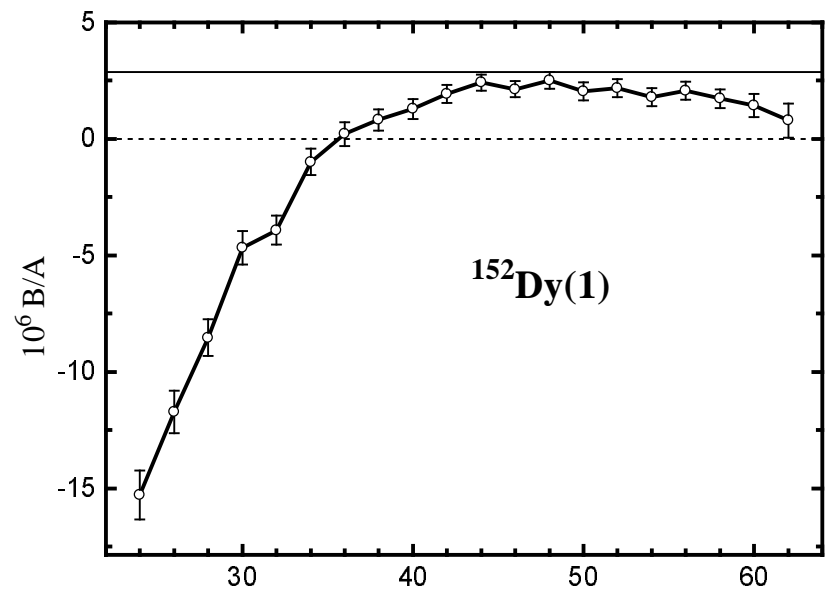
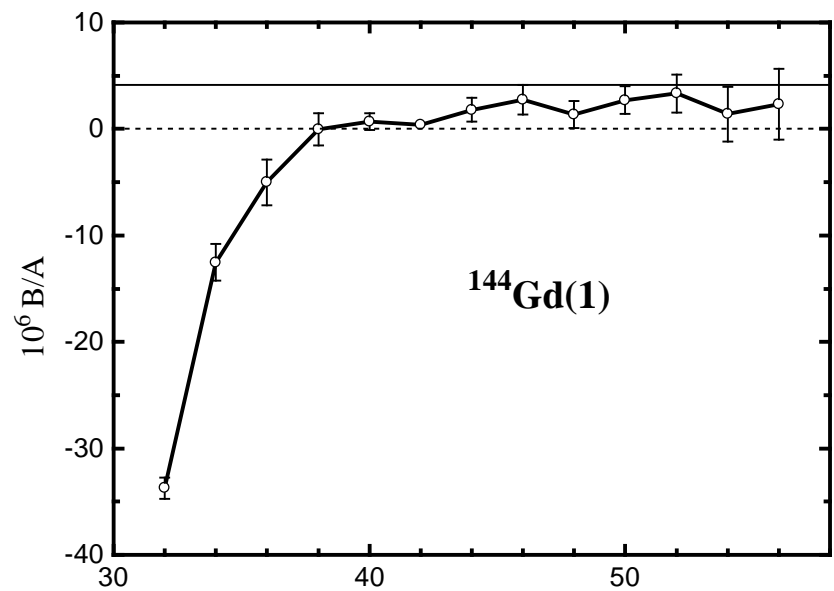
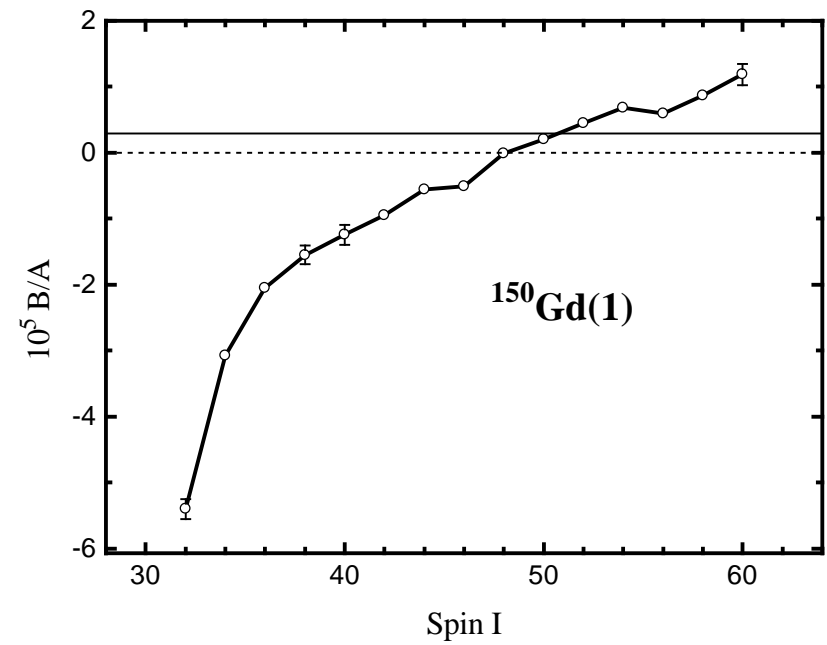
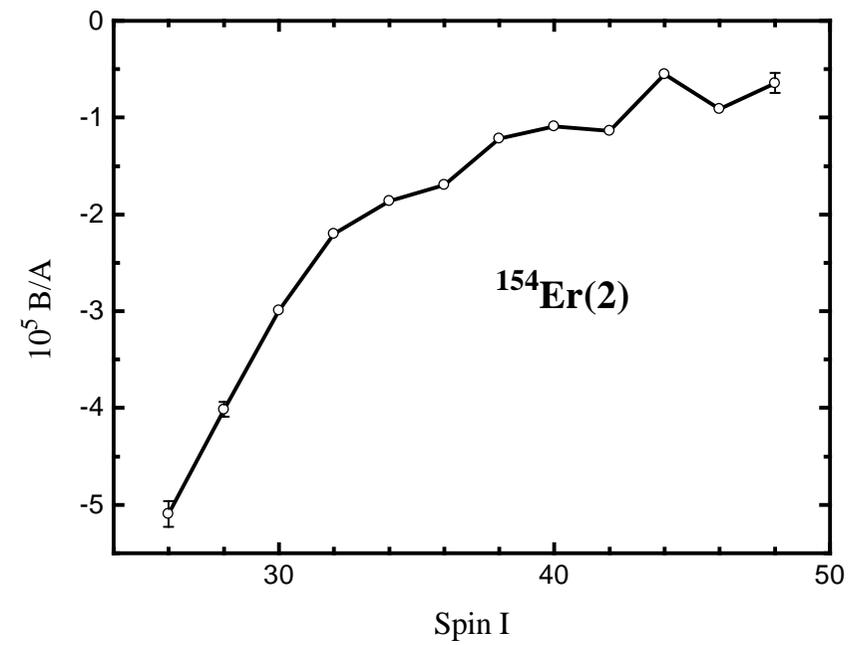
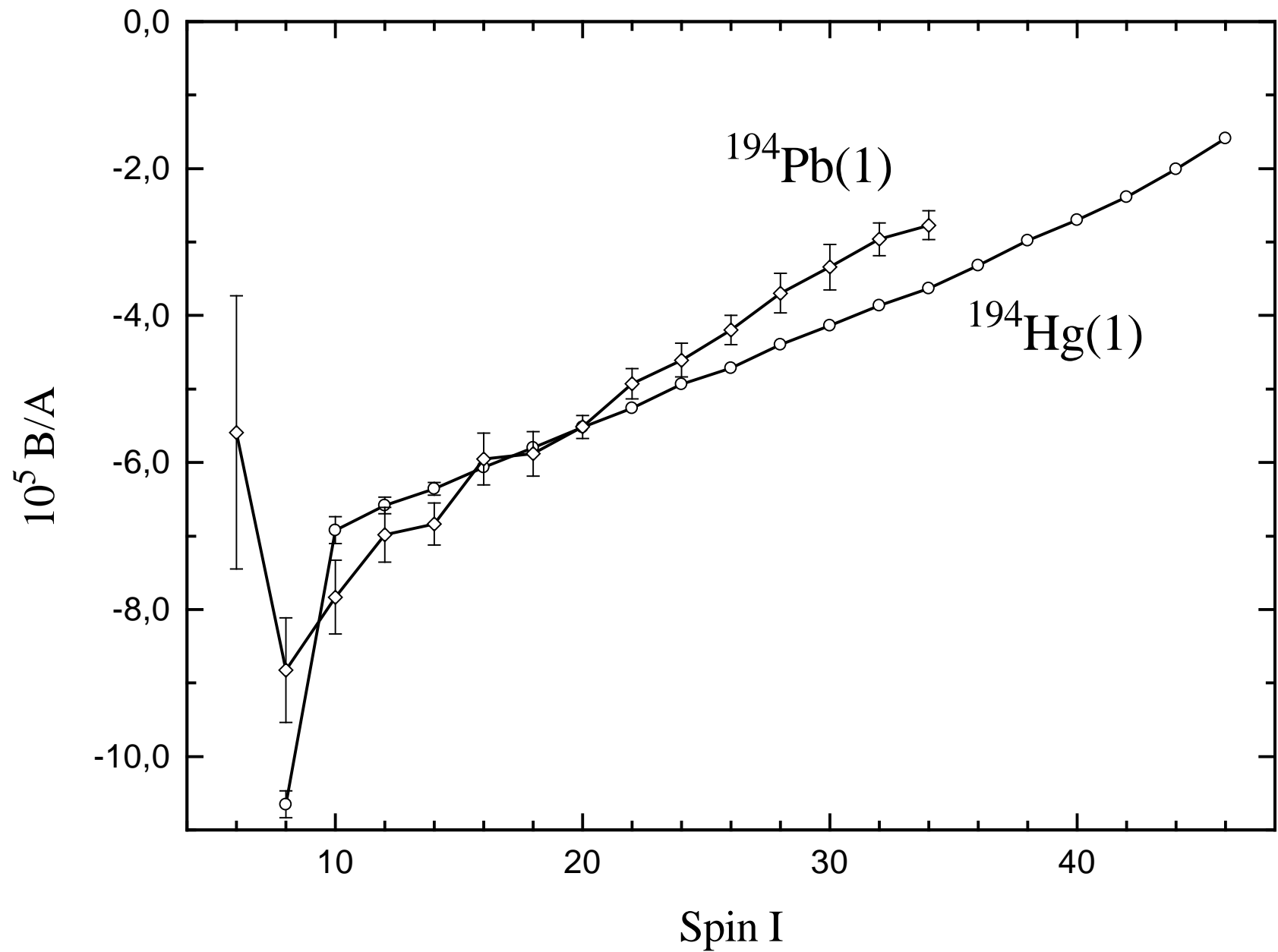
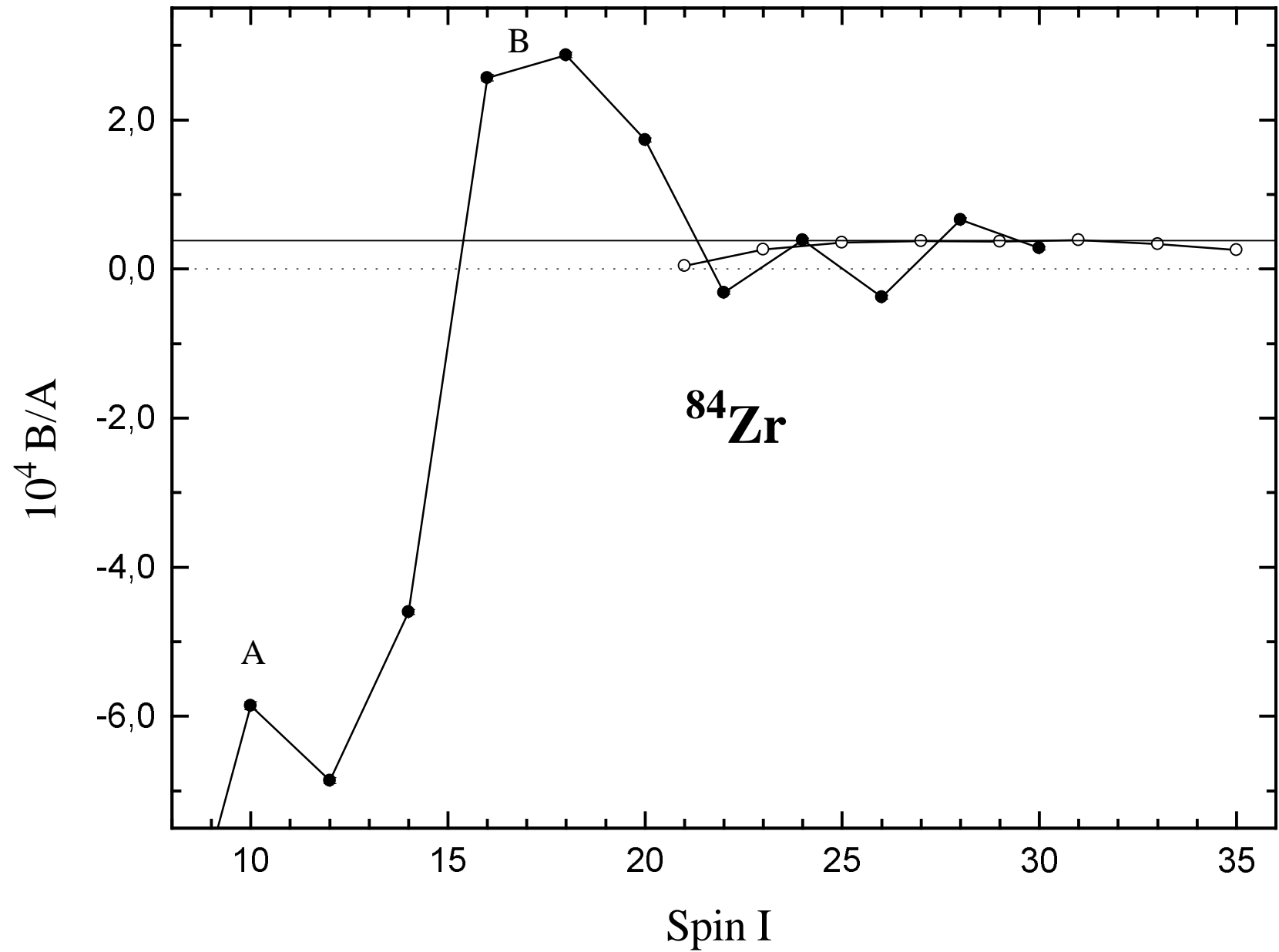
\end{document}